\documentclass[lettersize,onecolumn]{IEEEtran}
\IEEEoverridecommandlockouts

\usepackage{soul,times,amssymb,amsthm,amsmath,amsfonts,bbm,mathrsfs,xcolor,subfigure,graphicx,enumitem,booktabs,euscript,nicefrac}
\allowdisplaybreaks
\usepackage{graphicx}
\usepackage[unicode]{hyperref}
\hypersetup{hidelinks}
\usepackage{mathtools}
\usepackage[normalem]{ulem}
\usepackage[font=footnotesize]{caption}
\usepackage[utf8]{inputenc}
\usepackage[T1]{fontenc}
\usepackage{xargs}
\usepackage[colorinlistoftodos,prependcaption]{todonotes}

\usepackage{cleveref}
\crefname{equation}{}{}

\newtheorem{theorem}{Theorem}[section]
\newtheorem{lemma}[theorem]{Lemma}

\newtheorem{proposition}[theorem]{Proposition}
\newtheorem{corollary}[theorem]{Corollary}
\theoremstyle{remark}
\newtheorem{remark}{Remark}
\newtheorem{definition}{Definition}[section]
\newtheorem{assumption}[theorem]{Assumption}

\newcommand\nc\newcommand
\nc{\bfsl}{\bfseries\slshape}
\nc{\bfit}{\bfseries\itshape}
\nc{\sfsl}{\sffamily\slshape}
\nc{\dfn}{\sffamily\slshape\small}

\nc\bfa{{\boldsymbol a}}\nc\bfA{{\boldsymbol A}}\nc\cA{{\EuScript A}}
\nc\bfb{{\boldsymbol b}}\nc\bfB{{\boldsymbol B}}\nc\cB{{\EuScript B}}
\nc\bfc{{\boldsymbol c}}\nc\bfC{{\boldsymbol C}}\nc\cC{{\mathscr C}}
\nc\bfd{{\boldsymbol d}}\nc\bfD{{\boldsymbol D}}\nc\cD{{\EuScript D}}
\nc\bfe{{\boldsymbol e}}\nc\bfE{{\boldsymbol E}}\nc\cE{{\EuScript E}}
\nc\bff{{\boldsymbol f}}\nc\bfF{{\boldsymbol F}}\nc\cF{{\mathscr F}}
\nc\bfg{{\boldsymbol g}}\nc\bfG{{\boldsymbol G}}\nc\cG{{\EuScript G}}
\nc\bfh{{\boldsymbol h}}\nc\bfH{{\boldsymbol H}}\nc\cH{{\mathcal H}}
\nc\bfi{{\boldsymbol i}}\nc\bfI{{\boldsymbol I}}\nc\cI{{\mathcal I}}
\nc\bfj{{\boldsymbol j}}\nc\bfJ{{\boldsymbol J}}\nc\cJ{{\EuScript J}}
\nc\bfk{{\boldsymbol k}}\nc\bfK{{\boldsymbol K}}\nc\cK{{\EuScript K}}
\nc\bfl{{\boldsymbol l}}\nc\bfL{{\boldsymbol L}}\nc\cL{{\EuScript L}}
\nc\bfm{{\boldsymbol m}}\nc\bfM{{\boldsymbol M}}\nc\cM{{\EuScript M}}
\nc\bfn{{\boldsymbol n}}\nc\bfN{{\boldsymbol N}}\nc\cN{{\EuScript N}}
\nc\bfo{{\boldsymbol o}}\nc\bfO{{\boldsymbol O}}\nc\cO{{\EuScript O}}
\nc\bfp{{\boldsymbol p}}\nc\bfP{{\boldsymbol P}}\nc\cP{{\EuScript P}}
\nc\bfq{{\boldsymbol q}}\nc\bfQ{{\boldsymbol Q}}\nc\cQ{{\EuScript Q}}
\nc\bfr{{\boldsymbol r}}\nc\bfR{{\boldsymbol R}}\nc\cR{{\EuScript R}}
\nc\bfs{{\boldsymbol s}}\nc\bfS{{\boldsymbol S}}\nc\cS{{\EuScript S}}
\nc\bft{{\boldsymbol t}}\nc\bfT{{\boldsymbol T}}\nc\cT{{\EuScript T}}
\nc\bfu{{\boldsymbol u}}\nc\bfU{{\boldsymbol U}}\nc\cU{{\EuScript U}}
\nc\bfv{{\boldsymbol v}}\nc\bfV{{\boldsymbol V}}\nc\cV{{\mathscr V}}
\nc\bfw{{\boldsymbol w}}\nc\bfW{{\boldsymbol W}}\nc\cW{{\mathscr W}}
\nc\bfx{{\boldsymbol x}}\nc\bfX{{\boldsymbol X}}\nc\cX{{\EuScript X}}
\nc\bfy{{\boldsymbol y}}\nc\bfY{{\boldsymbol Y}}\nc\cY{{\mathscr Y}}
\nc\bfz{{\boldsymbol z}}\nc\bfZ{{\boldsymbol Z}}\nc\cZ{{\EuScript Z}}

\nc{\1}{{\mathbbm{1}}}
\nc\rr{{\mathbb R}}
\nc\ee{{\mathbb E}}
\nc\mS{{\mathcal S}}
\nc\mC{{\mathcal C}}
\nc\mD{{\mathcal D}}
\nc\mB{{\mathcal B}}
\nc\mT{{\mathcal T}}
\nc\mM{{\mathcal M}}
\nc\sB{{\mathfrak B}}
\nc{\integers}{{\mathbb Z}}
\nc{\ff}{{\mathbb F}}
\nc{\ii}{{\mathbb I}}
\nc{\sC}{{\mathfrak C}}
\nc{\sL}{{\mathfrak L}}
\nc\hH{{\mathsf H}}
\nc\gG{{\mathsf G}}
\nc{\remove}[1]{}
\nc\Renyi{R{\'e}nyi }
\nc\tbt{\textbullet\quad}
\nc\ffm{\ff_2[x_1, x_2, \dots, x_m]}

\DeclareSymbolFont{bbold}{U}{bbold}{m}{n}
\DeclareSymbolFontAlphabet{\mathbbold}{bbold}

\DeclareMathOperator{\supp}{supp}

\nc{\genstirlingI}[3]{ \genfrac{[}{]}{0pt}{#1}{#2}{#3}}
\nc{\genstirlingII}[3]{\genfrac{\{}{\}}{0pt}{#1}{#2}{#3}}
\nc{\stirlingI}[2]{\genstirlingI{}{#1}{#2}}
\nc{\dstirlingI}[2]{\genstirlingI{0}{#1}{#2}}
\nc{\tstirlingI}[2]{\genstirlingI{1}{#1}{#2}}
\nc{\stirlingII}[2]{\genstirlingII{}{#1}{#2}}
\nc{\dstirlingII}[2]{\genstirlingII{0}{#1}{#2}}
\nc{\tstirlingII}[2]{\genstirlingII{1}{#1}{#2}}

\nc\todoin[2][]{\todo[inline, caption={2do}, #1]{
		\begin{minipage}{\textwidth-4pt}#2\end{minipage}}}
\nc{\mynote}[2]{{\color{#1} \marginpar{\tiny #2}}}
\nc{\mybignote}[2]{{\color{#1} $\langle \langle$ #2$\rangle \rangle$}}
\newcommandx{\rednote}[2][1=]{\todo[linecolor=red,backgroundcolor=red!25,bordercolor=red,#1]{#2}}
\newcommandx{\bluenote}[2][1=]{\todo[linecolor=blue,backgroundcolor=blue!25,bordercolor=blue,#1]{#2}}
\newcommandx{\yellownote}[2][1=]{\todo[linecolor=yellow,backgroundcolor=yellow!25,bordercolor=yellow,#1]{#2}}
\newcommandx{\greennote}[2][1=]{\todo[inline,linecolor=olive,backgroundcolor=green!25,bordercolor=olive,#1]{#2}}

\nc{\new}[1]{{\color{blue} #1}}
\nc{\rmark}[1]{{\color{red} #1}}
\nc{\pmark}[1]{\textcolor[rgb]{0.50,0.00,1.0}{ #1}}
\nc{\brmark}[1]{\textcolor[rgb]{0.63,0.31,0.2}{ #1}}
\nc{\dgmark}[1]{\textcolor[rgb]{0.00,0.8,0.4}{ #1}}
\nc{\mmark}[1]{{\color{magenta} #1}}
\nc{\ab}[1]{{\footnotesize\color{red}{[AB: #1]}}}
\nc{\mad}[1]{{\footnotesize\color{blue}{[MP: #1]}}}
\nc{\madp}[1]{{\footnotesize\color{red}{[MP: #1]}}}
\nc{\sbh}[1]{{\footnotesize\color{red}{[SB: #1]}}}
\nc\redout{\bgroup\markoverwith{\textcolor{red}{\rule[0.5ex]{2pt}{0.8pt}}}\ULon}
\nc\blueout{\bgroup\markoverwith{\textcolor{blue}{\rule[0.5ex]{2pt}{0.8pt}}}\ULon}
\newcommand{\mc}{-\!\!\circ\!\!-}

\usepackage[backend=biber,style=ieee,sorting=none,doi=false,isbn=false,url=false,eprint=false]{biblatex}
\AtEveryBibitem{\clearfield{month}}
\AtEveryBibitem{\clearfield{day}}
\def\BibTeX{{\rm B\kern-.05em{\sc i\kern-.025em b}\kern-.08em
    T\kern-.1667em\lower.7ex\hbox{E}\kern-.125emX}}
\begin{document}
    \title{Tightness of the Partition Bound for Multiterminal Secret Key Capacity \\
    \thanks{$^\dag$ Madhura Pathegama was with the Department of Electrical and Computer Engineering and the Institute for Systems Research, University of Maryland, College Park, MD 20742. He is now with the School of Electrical and Computer Engineering at Georgia Institute of Technology, Atlanta, GA 30332. Email: macharige3@gatech.edu. 
    
    S. Bhattacharya was with the Department of Electrical and Computer Engineering and the Institute for Systems Research, University of Maryland, College Park, MD 20742, USA. He is now with Texas Instruments Inc, Dallas, TX 75243. Email: {sagnikbh96}@gmail.com. His research was supported by the U.S. National Science Foundation under Grant CCF 1910497 and Grant CCF 2310203.}
    }

    \author{\IEEEauthorblockN{M. Pathegama and S. Bhattacharya$^\dag$}}
    \maketitle

    \begin{abstract}
        We consider secret key agreement among \(m\) terminals, each
        observing one component of a finite-alphabet discrete memoryless
        multiple source. The terminals communicate over a noiseless public
        channel, and a target set \(A\subseteq[m]\) seeks to generate a
        shared secret key, with the remaining terminals acting as helpers.
        The partition-based Csisz\'ar--Narayan upper bound on secret key
        capacity is known to be tight without helpers but can be strict
        in their presence. We give a complete characterization of its tightness: 
        for \(2\leq |A|\leq m\), the bound is tight for every source if and only if
        \(|A|\in\{2,m-2,m-1,m\}\). To prove tightness, we use the
        supermodularity of conditional entropy and the structure of
        optimal supports in the linear program for communication for
        omniscience. We first establish the result under strict
        supermodularity and then remove this assumption by a
        perturbation argument. For all remaining pairs \((m,|A|)\),
        we extend Chan's six-terminal counterexample. We also study
        optimal partitions when there are two target terminals and
        the source is a Markov random field on a connected graph.
        We show that an optimal bipartition separating the targets
        can be chosen so that both blocks induce connected subgraphs.
        For cycles, this reduces the number of candidate bipartitions
        from exponential to at most quadratic in \(m\).
    \end{abstract}
        

    \section{Introduction}\label{sec:introduction}
        Let \([m]=\{1,\ldots,m\}\) be a set of \(m\geq 2\) terminals, and consider a discrete memoryless multiple source (DMMS) with component random variables (rvs) \(X_{[m]}=(X_1,\ldots,X_m)\), taking values in finite alphabets and having a known joint probability mass function (pmf) \(P_{X_{[m]}}\).
        In the \(m\)-terminal secret key agreement problem, terminal \(i\in[m]\) observes a sequence \(X_i^n=(X_{i1},\ldots,X_{in})\) of \(n\) independent and identically distributed (i.i.d.) realizations of the \(i\)th component of the DMMS.
        The terminals can communicate interactively over a noiseless, authenticated public broadcast channel, with the goal of enabling a subset \(A\subseteq[m]\) of ``target'' terminals, where \(2\leq |A|\leq m\), to agree on a shared secret key.
        The key is an rv that these terminals can agree on with high probability and that is nearly uniform and nearly independent of the public communication, which may be observed by an eavesdropper.
        The maximum achievable rate of such a secret key is called the \emph{secret key capacity} and is denoted by \(C_{\mathrm{SK}}(A)\).
        A central goal in the study of multiterminal secret key agreement is to characterize this capacity and construct protocols attaining it.
        The terminals in \(A^c=[m]\setminus A\) are called ``helpers.''
        The eavesdropper has no side information about the source beyond the public communication.
        Secrecy from the helpers is not required.

        The secret key agreement problem was first studied for the two-terminal case \(|A|=m=2\) by Maurer \cite{maurer1993secret} and Ahlswede and Csisz\'ar \cite{ahlswede1993common}.
        Csisz\'ar and Narayan \cite{csiszar2004secrecy} generalized the problem to arbitrary \(m\) and \(A\), establishing a close connection with the problem of \emph{communication for omniscience} (CO).
        In the latter problem, each terminal \(i\in A\) must reconstruct \(X_{[m]}^n=(X_1^n,\ldots,X_m^n)\) from its local observation \(X_i^n\) and interactive public communication, with error probability tending to zero as \(n\to\infty\).
        The smallest total communication rate required is called the \emph{minimum CO rate} and is denoted as \(R_{\mathrm{CO}}(A)\).
        They showed that
        \begin{align}
            C_{\mathrm{SK}}(A)+R_{\mathrm{CO}}(A)=H(X_{[m]}).
            \label{eq:csk-rco-sum}
        \end{align}
        Thus, determining the secret key capacity is equivalent to determining the minimum CO rate.
        
        Exploiting the fact that \(R_{\mathrm{CO}}(A)\) can be computed using a linear program (LP) based on single-letter Slepian--Wolf constraints, Csisz\'ar and Narayan expressed this rate as the maximum of a weighted sum of conditional entropies \cite{csiszar2004secrecy}; see \eqref{eq:rco-lp}. Each conditional entropy measures the uncertainty in the observations of one group of terminals given those of the remaining terminals. The optimization chooses how much weight to assign to each term, subject to constraints depending on the target set \(A\)\footnote{These admissible assignments of weights are called \emph{fractional partitions}; see \Cref{def:fractional-partitions}.}. Together with the identity above, this gives an exact single-letter characterization of the secret key capacity, but one that requires optimizing weights associated with many, possibly overlapping, groups of terminals.
        
        Csisz\'ar and Narayan also obtained a simpler lower bound for \(R_{\mathrm{CO}}(A)\) by restricting the allowed weights to those prescribed by ordinary partitions of \([m]\); see \eqref{eq:partition-rco-bound}. Here, the terminals are divided into disjoint groups, each containing at least one target terminal. Once the partition is chosen, both the conditional entropy terms and their weights are fixed. Thus, to obtain this lower bound, instead of optimizing the weights individually, one optimizes over these partitions. Equivalently, using \Cref{eq:csk-rco-sum}, this gives an upper bound for \(C_{\mathrm{SK}}(A)\), which can be written in terms of Kullback--Leibler (KL) divergence, as in \eqref{eq:upper-bound}. We shall henceforth refer to this upper bound as the \emph{partition upper bound}.
        
        The partition upper bound was shown to be tight in some special cases, e.g., when \(|A|=m=3\), and for general \(m\) and \(A\) when the component rvs form a Markov chain or, more generally, a Markov chain on a tree \cite{csiszar2004secrecy}. It was also later shown to be tight for a class of DMMSs arising from pairwise independent networks (PINs) when \(A=[m]\) and when \(|A|=2\), using arguments from graph theory \cite{nitinawarat2010secret,nitinawarat2010perfect}.
        
        In a significant advance, the partition upper bound was shown to be tight for \emph{all} \(m\) when there are no helpers, i.e., \(A=[m]\), in \cite{chan2008tightness,chan2010mutual}, using the submodularity of entropy and the linear program for \(R_{\mathrm{CO}}(A)\). In this case, \(C_{\mathrm{SK}}(A)\) equals \emph{shared information}, a measure of mutual dependence among multiple rvs that generalizes Shannon's mutual information to \(m\geq3\) rvs \cite{chan2015multivariate}. The underlying optimization over partitions defining shared information turns out to have a rich combinatorial structure that is useful in applications. Subsequent work has studied this structure and its implications for constructing omniscience protocols and for the communication requirements of secret key generation \cite{chan2015multivariate,chan2016successive,mukherjee2016public}.
        
        In the presence of helpers, the situation is far less satisfactory. It is known, using an example with \(m=6\) and \(|A|=3\), that the partition upper bound is \emph{not} tight in general \cite{chan2008tightness}. The known special cases and this counterexample, however, do not provide a general characterization of when the bound is tight in the presence of helpers.
        
        This is the main question we address in this paper: \emph{for which pairs \((m,|A|)\) is the partition upper bound tight for every \(m\)-terminal DMMS?} Equivalently, for what pairs \((m,|A|)\) can the optimization over weights always be replaced by an optimization over ordinary partitions, without changing the optimal value?
        
        Our results, together with the known no-helper case, give a complete characterization: the partition upper bound is tight for every DMMS when \(|A|\in\{2,m-2,m-1,m\}\), and is not tight in general for any other pair \((m,|A|)\). The \(|A|=2\) case generalizes the already known tightness in PIN models. Thus, among problems with at least three target terminals, the partition characterization remains universally valid with zero, one, or two helpers, but can fail once three helpers are present. The two-target case forms an additional exception.
        
        The proof is based on the LP characterization of \(R_{\mathrm{CO}}(A)\) and an equivalent formulation of the partition bound in terms of conditional entropies. Using supermodularity of $f(B) = H(X_B \,|\, X_{B^c})$ and properties of LPs, we show that, in the positive cases above, there is always an optimal choice of weights prescribed by an ordinary partition of \([m]\). We first prove this when $f(B)$ is \emph{strictly} supermodular. Then, for a general source, we suitably augment the terminals' observations with auxiliary random variables of arbitrarily small total entropy. Doing so makes an arbitrarily small change to the minimum CO rate and allows us to use our result for the strictly supermodular case; taking a limit then gives the desired result for the original source. For all remaining target-set sizes, we extend the six-terminal counterexample from \cite{chan2008tightness} by duplicating observations across suitable groups of terminals.        
        
        In addition to our general results concerning the tightness
        of the partition upper bound, we investigate the structure
        of optimal partitions when \(|A|=2\) and the source is a Markov
        random field on a connected graph. In this setting, the capacity
        is given by the minimum mutual information across bipartitions
        that separate the two target terminals and whose blocks induce
        connected subgraphs; see \Cref{thm:partition-for-A-2}.
        
        This result gives characterizations for particular graph families.
        For trees, it recovers the known expression as the minimum mutual
        information over the edges on the path connecting the targets
        \cite[Example~7]{csiszar2004secrecy}. For cycles, it suffices to
        consider bipartitions obtained by deleting one edge from each
        of the two paths connecting the targets. 
        The number of candidate
        bipartitions is then at most quadratic in \(m\), rather than
        exponential without
        the connectedness restriction.

        
        \subsection*{Layout of this paper}
        
        We begin with preliminaries in \Cref{sec:prelims} and state our main
        results in \Cref{sec:main}. Our proofs under strict supermodularity
        follow in \Cref{sec:proofs}, and the limiting argument in
        \Cref{sec:non-strict-supermodularity} extends them to the general
        case. In \Cref{sec:counterexample}, we extend Chan's counterexample
        to all remaining pairs \((m,|A|)\).
        In \Cref{sec:structural-A-2}, we study optimal
        bipartitions for Markov random fields with two target terminals
        and give the resulting characterizations for trees and cycles. 
        Finally, in \Cref{sec:conclusion}, we discuss extensions of our
results to settings with additional side information available to the terminals and the eavesdropper, as well as to jointly Gaussian sources, and highlight an open question concerning optimal partitions.

    \section{Preliminaries}\label{sec:prelims}
        We use \([m]\) to denote the set \(\{1,\ldots,m\}\). We will use upper-case letters towards the beginning of the alphabet, \(A,B,C\), to denote subsets of \([m]\), and their script versions, \(\mathcal A,\mathcal B,\mathcal C\), to denote collections of subsets. We will also use upper-case letters like \(X,Y,Z\) to denote rvs, and their script versions, \(\mathcal X,\mathcal Y,\mathcal Z\), to denote the respective alphabets. For $A \subseteq [m]$, let $A^c = [m] \setminus A$.
        
        \subsection*{Secret key agreement and communication for omniscience}
        
        Let \(X_1,\ldots,X_m\), \(m\geq2\), be rvs with finite alphabets \(\mathcal X_1,\ldots,\mathcal X_m\) and joint pmf \(P_{X_{[m]}}\). For \(B\subseteq[m]\), we write \(X_B=(X_i:i\in B)\), with alphabet \(\mathcal X_B=\prod_{i\in B}\mathcal X_i\), and take \(X_\emptyset\) to be constant. Throughout, the target set \(A\subseteq[m]\) satisfies \(2\leq |A|\leq m\).
        
        Consider a DMMS with generic rv \(X_{[m]}=(X_1,\ldots,X_m)\), constituted by \(n\) i.i.d.\ repetitions
        \[
            X_{[m]}^n
            =
            (X_{[m],1},\ldots,X_{[m],n}),
            \qquad
            X_{[m],t}=(X_{1t},\ldots,X_{mt}).
        \]
        We assume that each terminal \(i\in[m]\) observes \(X_i^n=(X_{i1},\ldots,X_{in})\). No private randomness other than $X_i^n$ is allowed at any terminal. All terminals are assumed to know \(P_{X_{[m]}}\).
        
        The terminals in \([m]\) engage in \(\ell=\ell_n\geq0\) rounds of interactive public broadcast communication, say in round-robin fashion, represented by
        \[
            F_1^j,\ldots,F_m^j,
            \qquad j=1,\ldots,\ell,
        \]
        with \(F_i^j\) being the communication in round \(j\) from terminal \(i\). Here, \(F_i^j\) is a function of \(X_i^n\), all communication from earlier rounds, and the preceding communication in round \(j\). Denote the entirety of the communication from all terminals by
        \[
            \mathbb F
            =
            \mathbb F^{(n)}
            =
            (F_1^1,\ldots,F_m^1,\ldots,F_1^\ell,\ldots,F_m^\ell),
        \]
        and let \(\mathcal F^{(n)}\) be the finite set of its possible values. When \(\ell=0\), $\mathbb{F}$ is constant.
        
        \begin{definition}[Secret key and secret key capacity]
            \label{def:secret-key}
            A sequence of rvs
            \(K^{(n)}=K^{(n)}(X_{[m]}^n)\), with values in finite
            sets \(\mathcal K^{(n)}\), constitutes a \emph{secret key} for
            the set of terminals \(A\) if there exists a sequence of public
            communication protocols for which rvs
            \[
                K_i^{(n)}
                =
                K_i^{(n)}(X_i^n,\mathbb F),
                \qquad i\in A,
            \]
            can be generated at the respective terminals so as to satisfy
            the \emph{key agreement} condition
            \begin{equation}\label{eq:key-agreement}
                \Pr\bigl(
                    K_i^{(n)}=K^{(n)}
                    \text{ for every }i\in A
                \bigr)
                \geq1-\epsilon_n
            \end{equation}
            and the \emph{secrecy} condition
            \begin{equation}\label{eq:divergence-secrecy}
                D\left(
                    P_{K^{(n)}\mathbb F}
                    \,\middle\|\,
                    P_{\mathrm{unif}\{\mathcal K^{(n)}\}}
                    \times P_{\mathbb F}
                \right)
                \leq\delta_n,
            \end{equation}
            where \(P_{\mathrm{unif}\{\mathcal K^{(n)}\}}\) is the uniform
            pmf on \(\mathcal K^{(n)}\), and
            \(\epsilon_n,\delta_n\to0\) as \(n\to\infty\).
        
            Furthermore, \(r\geq0\) is an \emph{achievable secret key rate}
            for \(A\) if there exists such a sequence of keys and protocols satisfying
            \[
                \liminf_{n\to\infty}
                \frac{1}{n}\log|\mathcal K^{(n)}|
                \geq r.
            \]
            The supremum of achievable secret key rates is the
            \emph{secret key capacity}, denoted by \(C_{\mathrm{SK}}(A)\),
            or by \(C_{\mathrm{SK}}(A;X_{[m]})\) when the dependence on
            the source needs to be explicit.
        \end{definition}
        
        \begin{definition}[Communication for omniscience]
            \label{def:comm-for-omniscience}
            A sequence of interactive communication protocols achieves
            \emph{omniscience} for the terminals in \(A\) if there exist
            decoding functions
            \[
                \phi_i^{(n)}:
                \mathcal X_i^n\times\mathcal F^{(n)}
                \longrightarrow
                \mathcal X_{[m]}^n,
                \qquad i\in A,
            \]
            such that
            \begin{equation}\label{eq:omniscience}
                \Pr\bigl(
                    \phi_i^{(n)}(X_i^n,\mathbb F^{(n)})
                    =X_{[m]}^n
                    \text{ for every }i\in A
                \bigr)
                \geq1-\epsilon_n,
            \end{equation}
            where \(\epsilon_n\to0\) as \(n\to\infty\).

            A number \(r\geq0\) is an \emph{achievable rate of communication
            for omniscience} (CO rate) if there exists such a sequence of protocols
            satisfying
            \[
                \limsup_{n\to\infty}
                \frac{1}{n}\log|\mathcal F^{(n)}|
                \leq r.
            \]
            The rate counts the entire public communication, including
            messages from helpers, although only the terminals in \(A\)
            are required to reconstruct \(X_{[m]}^n\).
        
            The infimum of achievable CO rates is denoted by \(R_{\mathrm{CO}}(A)\) or \(R_{\mathrm{CO}}(A;X_{[m]})\) with the dependence on the source variables made explicit.
        \end{definition}
        
        \subsection*{The LP characterization and the partition bound}
        
        To state the LP characterization of \(R_{\mathrm{CO}}(A)\), define
        \[
            \mathcal B(A)
            \coloneqq
            \{B\subseteq[m]:B\neq\emptyset,\ A\nsubseteq B\}.
        \]
        
        Thus, \(\mathcal B(A)\) consists of the nonempty subsets that omit at least one element of $A$.
        
        \begin{definition}[Fractional partitions]
            \label{def:fractional-partitions}
            A family of nonnegative weights
            \(\lambda=\{\lambda_B:B\in\mathcal B(A)\}\) satisfying
            \begin{equation}
                \sum_{\substack{B\in\mathcal B(A)\\i\in B}}\lambda_B=1
                \qquad\text{for every }i\in[m] \label{eq:balance}
            \end{equation}
            is called a \emph{fractional partition}. We denote the set
            of all such fractional partitions for a given \(A\) by
            \(\Lambda(A)\), and call the equality in \Cref{eq:balance} the
            \emph{balance condition} for \(i\).
        
            For \(\lambda\in\Lambda(A)\), its \emph{support} is
            \[
                \supp(\lambda)
                \coloneqq
                \{B\in\mathcal B(A):\lambda_B>0\}.
            \]
        \end{definition}
        
        The two rates \(C_{\mathrm{SK}}(A)\) and \(R_{\mathrm{CO}}(A)\) are
        intimately related.
        
        \begin{theorem}[Secret key capacity and communication for omniscience
            \cite{csiszar2004secrecy}]\label{thm:sk-capacity-omniscience}
            \begin{equation}\label{eq:si-rco-rel}
                C_{\mathrm{SK}}(A)+R_{\mathrm{CO}}(A)
                =
                H(X_{[m]}).
            \end{equation}
            Further, the communication rate for omniscience
            \(R_{\mathrm{CO}}(A)\) is the optimal value of the linear program
            \begin{equation}\label{eq:rco-lp}
                R_{\mathrm{CO}}(A)
                =
                \max_{\lambda\in\Lambda(A)}
                \sum_{B\in\mathcal B(A)}
                \lambda_B H(X_B\,|\,X_{B^c}).
            \end{equation}
        \end{theorem}
        
        \begin{remark}[Slepian--Wolf constraints and LP duality]
            Naturally, the noninteractive version of the omniscience problem
            can be formulated as a Slepian--Wolf coding problem. It was shown
            in \cite[Proposition~1]{csiszar2004secrecy} that the corresponding
            minimum sum-rate is given by the LP below, and that allowing
            interactive communication does not reduce this minimum rate:
            \begin{equation}\label{eq:rco-primal}
                \begin{aligned}
                    R_{\mathrm{CO}}(A)
                    =\min_{R_1,\dots,R_m\in\rr}\quad
                    &\sum_{i=1}^m R_i\\
                    \textup{subject to}\quad
                    &\sum_{i\in B}R_i
                    \geq H(X_B\,|\,X_{B^c}),
                    \qquad B\in\mathcal B(A).
                \end{aligned}
            \end{equation}
            The constraints above already enforce \(R_i\geq0\), so the variables
            may be treated as unrestricted when taking the dual. Strong LP
            duality then yields \eqref{eq:rco-lp}, with the balance conditions
            defining \(\Lambda(A)\).
            
            The equality \eqref{eq:si-rco-rel} is a separate result
            which implies that an optimal communication scheme for omniscience
            allows extraction of a secret key at the optimal rate
            \(C_{\mathrm{SK}}(A)\).
        \end{remark}
        
        As shown in \cite[Example~4]{csiszar2004secrecy}, we can use the
        linear program in \Cref{thm:sk-capacity-omniscience} to find lower
        bounds on \(R_{\mathrm{CO}}(A)\) and, consequently, upper bounds
        on \(C_{\mathrm{SK}}(A)\). The bounds involve special
        \emph{partitions} of \([m]\), captured in the following definition.
        
        \begin{definition}[Admissible partitions]\label{def:partitions}
            Given \(A\subseteq[m]\) and \(2\leq k\leq |A|\), let
            \(\Pi_k(A)\) consist of partitions
            \(\pi=\{\pi_1,\ldots,\pi_k\}\) of \([m]\) into \(k\) nonempty
            blocks (or atoms) such that
            \[
                \pi_i\cap A\neq\emptyset
                \qquad\text{for every }i\in[k].
            \]
            We write
            \[
                \Pi(A)
                \coloneqq
                \bigcup_{k=2}^{|A|}\Pi_k(A).
            \]
            Since every block intersects \(A\) and there are at least two
            blocks, no block contains all of \(A\). 
        \end{definition}
        
        For any \(\pi=\{\pi_1,\ldots,\pi_k\}\in\Pi_k(A)\), set
        \[
            \lambda_B^\pi
            =
            \begin{cases}
                \dfrac{1}{k-1},
                &B=\pi_i^c\text{ for some }i\in[k],\\[1ex]
                0,
                &\text{otherwise}.
            \end{cases}
        \]
        Each complement \(\pi_i^c\) belongs to \(\mathcal B(A)\), since
        \(\pi_i\cap A\neq\emptyset\). Moreover, \(\lambda^\pi\in\Lambda(A)\),
        since each terminal belongs to exactly \(k-1\) of the complements
        \(\pi_1^c,\ldots,\pi_k^c\). Restricting the maximization in
        \eqref{eq:rco-lp} to these fractional partitions therefore gives
        the lower bound
        \begin{equation}\label{eq:partition-rco-bound}
            R_{\mathrm{CO}}(A)
            \geq
            \max_{2\leq k\leq |A|}
            \max_{\pi\in\Pi_k(A)}
            \frac{1}{k-1}
            \sum_{i=1}^k H(X_{\pi_i^c}\,|\,X_{\pi_i}).
        \end{equation}
        
        To obtain the corresponding upper bound on secret key capacity,
        observe that, for any \(\pi\in\Pi_k(A)\),
        \begin{align}\label{eq:partition-identity}
            \frac{1}{k-1}
            \sum_{i=1}^k H(X_{\pi_i^c}\,|\,X_{\pi_i})
            &=
            H(X_{[m]})
            -
            \frac{\sum_{i=1}^k H(X_{\pi_i})-H(X_{[m]})}{k-1}
            \notag\\
            &=
            H(X_{[m]})
            -
            \frac{1}{k-1}
            D\left(
                P_{X_{[m]}}
                \,\middle\|\,
                \prod_{i=1}^k P_{X_{\pi_i}}
            \right).
        \end{align}
        Substituting this identity into \eqref{eq:partition-rco-bound}
        and using \eqref{eq:si-rco-rel} yields the
        partition upper bound
        \begin{equation}\label{eq:upper-bound}
            C_{\mathrm{SK}}(A)
            \leq
            \min_{2\leq k\leq |A|}
            \min_{\pi\in\Pi_k(A)}
            \frac{1}{k-1}
            D\left(
                P_{X_{[m]}}
                \,\middle\|\,
                \prod_{i=1}^k P_{X_{\pi_i}}
            \right).
        \end{equation}
        For \(A=[m]\), the quantity on the right is called
        \emph{shared information} and equality holds in \Cref{eq:upper-bound} \cite{chan2008tightness,chan2010mutual,chan2015multivariate}.
        
        \subsection*{Supermodularity}
        
        To prove our main result, we need the notion of a \emph{supermodular}
        set function.
        
        \begin{definition}[Supermodularity]
            Let \(\Omega\) be a finite set. A set function
            \(f:2^\Omega\to\rr\) is called \emph{supermodular} if, for
            every \(B,B'\subseteq\Omega\),
            \[
                f(B)+f(B')
                \leq
                f(B\cup B')+f(B\cap B'),
            \]
            and \emph{strictly supermodular} if this inequality is strict
            whenever \(B\) and \(B'\) are incomparable, that is, whenever
            \(B\nsubseteq B'\) and \(B'\nsubseteq B\).
            A set function \(f\) is called submodular, respectively
            strictly submodular, if \(-f\) is supermodular, respectively
            strictly supermodular.
        \end{definition}
        
        The set function used throughout our proofs is
        \begin{equation}\label{eq:conditional-entropy-set-function}
            f(B)
            \coloneqq
            H(X_B\,|\,X_{B^c})
            =
            H(X_{[m]})-H(X_{B^c}),
            \qquad B\subseteq[m].
        \end{equation}
        It is supermodular by the submodularity of entropy, but need not
        be strictly supermodular.

    \section{Main Results}\label{sec:main}
        We show tightness of \eqref{eq:upper-bound} for
        \(|A|=2,m-2,m-1\), and reprove tightness for \(|A|=m\). 
        Together with the counterexamples in \Cref{sec:counterexample},
        this gives the following characterization.
        
        \begin{theorem}[Tightness of the partition bound]
            \label{thm:SK-cap-main} 
            Fix \(m\geq2\) and \(A\subseteq[m]\) with \(2\leq |A|\leq m\).
            The equality
            \begin{align}
                C_{\mathrm{SK}}(A)
                =
                \min_{2\leq k\leq |A|}
                \min_{\pi\in\Pi_k(A)}
                \frac{1}{k-1}
                D\left(
                    P_{X_{[m]}}
                    \,\middle\|\,
                    \prod_{i=1}^k P_{X_{\pi_i}}
                \right) \label{eq:sk-capacity}
            \end{align}
            holds for every finite-alphabet DMMS \(X_{[m]}\) if and only if
            \[
                |A|\in\{2,m-2,m-1,m\}.
            \]
            For every remaining pair, namely \(3\leq |A|\leq m-3\), there
            exists a finite-alphabet DMMS for which the inequality in
            \eqref{eq:upper-bound} is strict.
        \end{theorem}
        
        We prove the positive cases of \Cref{thm:SK-cap-main} by first proving the corresponding statement for $R_\mathrm{CO}(A)$, which we state in \Cref{thm:main} below. Using \eqref{eq:partition-identity} and the relationship between the SK capacity and minimum CO rate described in \eqref{eq:si-rco-rel} 
         then leads to \Cref{thm:SK-cap-main}. The remaining cases are established in
        \Cref{sec:counterexample} by extending the six-terminal
        counterexample of \cite{chan2008tightness}.

        \begin{remark}
            Unlike the existing proof for the case $A = [m]$ in \cite{chan2008tightness,chan2010mutual} (see also \cite{chan2015multivariate}), that relies on the concept of Dilworth truncation from the theory of submodular optimization, our approach only uses supermodularity and simple manipulations of the LP \Cref{eq:rco-lp}. Therefore, our proof remains of interest even in the previously established $A = [m]$ case.
        \end{remark}
        
        To state the corresponding result for $R_\mathrm{CO}(A)$, we need
        the following definition.
        
        \begin{definition}[Matched elements and regular collections]
            \label{def:matched-regular}
            Let \(\mathcal C\subseteq\mathcal B(A)\) be a collection of
            subsets of $[m]$. For \(x,y\in[m]\), we say that \(x\) and \(y\) are
            \emph{matched} in \(\mathcal C\) if
            \[
                x\in C
                \quad\Longleftrightarrow\quad
                y\in C
                \qquad\text{for every }C\in\mathcal C.
            \]
            Further, the collection \(\mathcal C\) with $|\mathcal{C}| \geq 2$ is
            \emph{\(A\)-regular} if
            \begin{enumerate}[label=(\roman*)]
                \item Every \(x\in A\) is omitted from exactly one set
                in \(\mathcal C\).
                \item Every \(x\in A^c\) is matched in \(\mathcal C\)
                with some \(y\in A\).
            \end{enumerate}
        \end{definition}
        
        The following lemma shows that \(A\)-regular collections of subsets
        are closely related to the partitions in \Cref{def:partitions}. We use this fact later to represent the optimal $R_\mathrm{CO}(A)$ using $A$-regular collections.
        
        \begin{lemma}\label{lem:partition-A-regular}
            The map
            \[
                \{\pi_1,\ldots,\pi_k\}
                \longmapsto
                \{\pi_1^c,\ldots,\pi_k^c\}
            \]
            is a one-to-one correspondence between admissible partitions \(\Pi(A)\) and the
            \(A\)-regular collections contained in \(\mathcal B(A)\).
        \end{lemma}
        
        \begin{proof}
            Let \(\pi=\{\pi_1,\ldots,\pi_k\}\in\Pi(A)\). Then,
            \(\mathcal C \coloneqq \{\pi_1^c,\ldots,\pi_k^c\}\subseteq\mathcal B(A)\).
            Each terminal belongs to exactly one block of \(\pi\), so it is omitted
            from exactly one set in \(\mathcal C\). Moreover, every
            helper shares its partition block with a target terminal
            because every block intersects \(A\). These two terminals
            have identical membership patterns in \(\mathcal C\).
            Hence \(\mathcal C\) is \(A\)-regular.
        
            Conversely, let
            \(\mathcal C=\{C_1,\ldots,C_k\}\subseteq\mathcal B(A)\) be
            \(A\)-regular. Condition~(i) implies that every target
            terminal is omitted from exactly one \(C_i\), and
            condition~(ii) extends this property to every helper.
            Thus the complements \(C_1^c,\ldots,C_k^c\) are pairwise
            disjoint and cover \([m]\). Since \(C_i\in\mathcal B(A)\),
            each complement is nonempty, proper, and intersects \(A\).
            Thus \(k\geq2\), while disjointness and the fact that every
            complement intersects \(A\) imply \(k\leq|A|\). Therefore,
            \(\{C_1^c,\ldots,C_k^c\}\in\Pi(A)\). Taking complements
            proves the correspondence.
        \end{proof}
        
        The next lemma shows that every \(A\)-regular collection supports
        a feasible fractional partition, and that its nonzero weights
        are uniquely determined by the balance conditions \Cref{eq:balance}.
        
        \begin{lemma}\label{lem:a-regular-supports-lambda}
            Every \(A\)-regular collection
            \(\mathcal C\subseteq\mathcal B(A)\) is the support of a unique
            fractional partition \(\lambda\in\Lambda(A)\), given by
            \[
                \lambda_B
                =
                \begin{cases}
                    \dfrac{1}{|\mathcal C|-1},
                    &B\in\mathcal C,\\[1ex]
                    0,
                    &B\notin\mathcal C.
                \end{cases}
            \]
        \end{lemma}
        
        \begin{proof}
            Write \(k=|\mathcal C|\). By
            \Cref{lem:partition-A-regular}, \(k\geq2\), and every terminal
            belongs to exactly \(k-1\) sets in \(\mathcal C\). Therefore,
            assigning weight \(1/(k-1)\) to each set in \(\mathcal C\)
            satisfies every balance condition \Cref{eq:balance}.
        
            Conversely, suppose that \(\lambda\in\Lambda(A)\) has support
            \(\mathcal C\), and write
            \[
                W\coloneqq\sum_{B\in\mathcal C}\lambda_B.
            \]
            For any \(C\in\mathcal C\), choose \(x\in A\setminus C\),
            which exists because \(C\in\mathcal B(A)\). By
            \(A\)-regularity, \(C\) is the only set in \(\mathcal C\)
            omitting \(x\). The balance condition \Cref{eq:balance} for \(x\) therefore
            gives
            \[
                W-\lambda_C=1.
            \]
            Hence all weights on \(\mathcal C\) equal \(W-1\).
            Summing over \(\mathcal C\) yields \(W=k(W-1)\), and thus
            \(\lambda_C=1/(k-1)\) for every \(C\in\mathcal C\).
        \end{proof}
        
        We can now state the corresponding result for the omniscience
        rate. We show that the maximization over fractional partitions
        in \eqref{eq:rco-lp} can be restricted to those fractional
        partitions that have \(A\)-regular supports. Since SK capacity and minimum CO rate are related as in \eqref{eq:si-rco-rel} and \eqref{eq:partition-identity}, proving \Cref{thm:main} is equivalent to proving the corresponding lower bound to \Cref{eq:upper-bound}, leading to the positive cases of \Cref{thm:SK-cap-main}.

        \begin{theorem}[Partition characterization of the minimum CO rate]
            \label{thm:main}
            When \(|A|\in\{2,m-2,m-1,m\}\), the LP \eqref{eq:rco-lp}
            has an optimal solution whose support is \(A\)-regular.
            Consequently,
            \begin{align}
                R_{\mathrm{CO}}(A)
                &=
                \max_{\pi\in\Pi(A)}
                \frac{1}{|\pi|-1}
                \sum_{D\in\pi}H(X_{D^c}\,|\,X_D)\nonumber\\
                &=
                \max_{\substack{
                    \mathcal C\subseteq\mathcal B(A)\\
                    \mathcal C\text{ is }A\text{-regular}
                }}
                \frac{1}{|\mathcal C|-1}
                \sum_{C\in\mathcal C}H(X_C\,|\,X_{C^c}).\label{eq: reg}
            \end{align}
        \end{theorem}
        
        By \Cref{lem:a-regular-supports-lambda}, every \(A\)-regular
        collection supports a feasible fractional partition with equal
        nonzero weights. Thus, to prove \Cref{thm:main}, it remains to
        show that an optimal solution of the LP \eqref{eq:rco-lp} can be chosen
        with an \(A\)-regular support. We prove this assuming strict
        supermodularity of $f(B) = H(X_B|X_{B^c})$ (see \Cref{eq:conditional-entropy-set-function}) in \Cref{sec:proofs}, and then remove the
        assumption in \Cref{sec:non-strict-supermodularity} by perturbing the joint pmf $P_{X_{[m]}}$ followed by a limiting argument.


    \section{Proofs assuming strict supermodularity}\label{sec:proofs}
        In this section, we prove \Cref{thm:main} for
        \(|A|=m,m-1,m-2\), and \(2\), under the following assumption.
        
        \begin{assumption}[Strict supermodularity]
            \label{ass:strict-supermodularity}
            The set function
            \[
                B\longmapsto H(X_B\,|\,X_{B^c}),
                \qquad B\subseteq[m],
            \]
            is strictly supermodular. That is, for any
            \(B,B'\subseteq[m]\) with \(B\nsubseteq B'\) and
            \(B'\nsubseteq B\),
            \begin{align*}
                &H(X_B\,|\,X_{B^c})
                +H(X_{B'}\,|\,X_{(B')^c}) <
                H(X_{B\cup B'}\,|\,X_{(B\cup B')^c})
                +H(X_{B\cap B'}\,|\,X_{(B\cap B')^c}).
            \end{align*}
        \end{assumption}
        
        This assumption is in force throughout this section and will be
        removed in \Cref{sec:non-strict-supermodularity}. Strict
        supermodularity imposes additional structure on the supports of
        optimal solutions, as described by the covering property in
        \Cref{lem:optimal-incomparable-cover}. This simplifies the proof
        by avoiding some of the difficulties that arise for general joint
        distributions. Once the result is established under this assumption,
        a straightforward limiting argument extends it to the general case.
        
        \subsection{Properties of feasible, optimal and extreme solutions}
        
        We begin with two basic properties that hold for every feasible
        solution of the LP \eqref{eq:rco-lp}, independently of the strict-supermodularity assumption. We will later combine these with supermodularity to derive structural properties of optimal solutions to \eqref{eq:rco-lp}.
        
        \begin{lemma}\label{lem:support-of-feasible-sol}
            Let \(\lambda\in\Lambda(A)\), and let
            \(\mathcal C=\supp(\lambda)\). Then:
            \begin{enumerate}[label=(\roman*)]
                \item Every \(C\in\mathcal C\) satisfies \(A\nsubseteq C\).
                \item For every \(x\in[m]\), there exists \(C\in\mathcal C\)
                such that \(x\notin C\).
            \end{enumerate}
        \end{lemma}
        
        \begin{proof}
            Part~(i) follows from
            \(\mathcal C\subseteq\mathcal B(A)\).
        
            To prove (ii), suppose that some \(x\in[m]\) belongs to every
            set in \(\mathcal C\). The balance condition
            \eqref{eq:balance} for \(x\) then gives
            \[
                \sum_{C\in\mathcal C}\lambda_C=1.
            \]
            Choose any \(C_0\in\mathcal C\) and \(y\in A\setminus C_0\),
            which exists by (i). Since \(\lambda_{C_0}>0\),
            \[
                \sum_{\substack{C\in\mathcal C\\y\in C}}\lambda_C
                \leq
                \sum_{C\in\mathcal C}\lambda_C-\lambda_{C_0}
                <1,
            \]
            contradicting the balance condition for \(y\).
        \end{proof}

        Optimality imposes an additional restriction on the support.
        To state it, we introduce the following terminology.
        
        \begin{definition}[Comparable sets]\label{def:comparable-sets}
            Two sets \(C,C'\subseteq[m]\) are \emph{comparable}, written
            \(C\sim C'\), if \(C\subseteq C'\) or \(C'\subseteq C\).
            Otherwise, they are \emph{incomparable}, and we write
            \(C\nsim C'\).
        \end{definition}
        
        \begin{lemma}[Covering property]
            \label{lem:optimal-incomparable-cover}
            Under Assumption~\ref{ass:strict-supermodularity}, let
            \(\lambda\in\Lambda(A)\) be an optimal solution to
            \eqref{eq:rco-lp}, and let \(\mathcal C=\supp(\lambda)\).
            If \(C,C'\in\mathcal C\) are incomparable, then
            \[
                A\subseteq C\cup C'.
            \]
        \end{lemma}
        
        \begin{proof}
            Suppose, for contradiction, that \(A\nsubseteq C\cup C'\).
            Then \(C\cup C'\in\mathcal B(A)\), and
            \(C\cap C'\in\mathcal B(A)\) whenever the intersection is
            nonempty. We construct a feasible fractional partition
            \(\tilde\lambda\) with a strictly larger objective value.
        
            Choose \(0<\epsilon<\min\{\lambda_C,\lambda_{C'}\}\).
            If \(C\cap C'\neq\emptyset\), define
            \[
                \begin{aligned}
                    \tilde\lambda_C
                    &=\lambda_C-\epsilon,
                    &\tilde\lambda_{C'}
                    &=\lambda_{C'}-\epsilon,\\
                    \tilde\lambda_{C\cup C'}
                    &=\lambda_{C\cup C'}+\epsilon,
                    &\tilde\lambda_{C\cap C'}
                    &=\lambda_{C\cap C'}+\epsilon.
                \end{aligned}
            \]
            If \(C\cap C'=\emptyset\), there is no weight associated
            with the intersection, and we instead set
            \[
                \tilde\lambda_C=\lambda_C-\epsilon,
                \qquad
                \tilde\lambda_{C'}=\lambda_{C'}-\epsilon,
                \qquad
                \tilde\lambda_{C\cup C'}=\lambda_{C\cup C'}+\epsilon.
            \]
            In either case, leave all other weights unchanged.
        
            The choice of \(\epsilon\) ensures that the new weights
            are nonnegative. Moreover, the identity
            \[
                \mathbf{1}_C+\mathbf{1}_{C'}
                =
                \mathbf{1}_{C\cup C'}+\mathbf{1}_{C\cap C'},
            \]
            where \(\mathbf{1}_D\) denotes the indicator vector of
            \(D\), shows that every balance condition \Cref{eq:balance} is preserved.
            This also covers the empty-intersection case, since
            \(\mathbf{1}_{\emptyset}=0\). Thus
            \(\tilde\lambda\in\Lambda(A)\).
        
            Since \(H(X_\emptyset\,|\,X_{[m]})=0\), the change in the
            objective in either case is
            \begin{align*}
                &\sum_{B\in\mathcal B(A)}
                (\tilde\lambda_B-\lambda_B)H(X_B\,|\,X_{B^c})\\
                &\quad=
                \epsilon\Bigl[
                    H(X_{C\cup C'}\,|\,X_{(C\cup C')^c})
                    +H(X_{C\cap C'}\,|\,X_{(C\cap C')^c})
                    -H(X_C\,|\,X_{C^c})
                    -H(X_{C'}\,|\,X_{(C')^c})
                \Bigr]
                >0,
            \end{align*}
            where the strict inequality follows from
            Assumption~\ref{ass:strict-supermodularity}.
            This contradicts the optimality of \(\lambda\).
        \end{proof}

        \begin{remark}
            The covering property in \Cref{lem:optimal-incomparable-cover} will be used repeatedly in our arguments, and it crucially relies on strict supermodularity. Furthermore, strict supermodularity is used in the arguments \emph{only} via \Cref{lem:optimal-incomparable-cover}.
        \end{remark}

        The feasible set of \eqref{eq:rco-lp}, $\Lambda(A)$, 
        is nonempty since assigning weight one to every singleton and
        zero to all other sets gives a feasible solution. It is also closed and bounded because of the balance condition \Cref{eq:balance}. Therefore, the LP described in \eqref{eq:rco-lp} attains its maximum at an \emph{extreme} solution, i.e., a solution that cannot
        be expressed as a nontrivial convex combination of two distinct
        feasible solutions. We next record a basic property of extreme solutions.

        \begin{lemma}[Support of an extreme solution is minimal]
            \label{lem:extreme-support}
            Let $\mu\in\Lambda(A)$ be an extreme solution to \eqref{eq:rco-lp}. Then $\supp(\mu)$ is minimal among the supports of elements of $\Lambda(A)$. i.e., for every $\lambda\in\Lambda(A)$,
            \[
                \supp(\lambda)\subseteq\supp(\mu)
                \quad\Longrightarrow\quad
                \supp(\lambda)=\supp(\mu).
            \]
            \end{lemma}
        
        \begin{proof}
            Suppose, for contradiction, that there exists
            \(\lambda\in\Lambda(A)\) such that
            \[
                \supp(\lambda)\subseteq\supp(\mu)
                \, \text{and}\
                \supp(\lambda)\neq\supp(\mu),
            \]
            i.e., $\supp(\lambda)\subsetneq\supp(\mu)$.
 
            Choose \(\alpha\in(0,1)\) sufficiently small that
            \(\alpha\lambda_B<\mu_B\) for every
            \(B\in\supp(\lambda)\), and define
            \[
                \lambda'_B
                \coloneqq
                \frac{\mu_B-\alpha\lambda_B}{1-\alpha},
                \qquad B\in\mathcal B(A).
            \]
            The weights \(\lambda'_B\) are nonnegative, and the balance
            conditions \Cref{eq:balance} for \(\lambda\) and \(\mu\) give
            \[
                \sum_{\substack{B\in\mathcal B(A)\\i\in B}}\lambda'_B
                =
                \frac{1-\alpha}{1-\alpha}
                =
                1
                \qquad\text{for every }i\in[m].
            \]
            Thus \(\lambda'\in\Lambda(A)\), and
            \[
                \mu=\alpha\lambda+(1-\alpha)\lambda'.
            \]
            
            The strict inclusion of supports gives \(\lambda\neq\mu\),
    and hence \(\lambda'\neq\lambda\). Thus this is a nontrivial
    convex decomposition of \(\mu\), contradicting its extremality.
        \end{proof}
        
Our main technical result in this section is the following.

\begin{proposition}\label{prop:extreme-optimal-regular}
    Suppose Assumption~\ref{ass:strict-supermodularity} holds and
    \(|A|\in\{2,m-2,m-1,m\}\). Then every extreme optimal solution
    to \eqref{eq:rco-lp} has an \(A\)-regular support.
\end{proposition}

It immediately yields \Cref{thm:main} under Assumption~\ref{ass:strict-supermodularity}.

\begin{proof}[Proof of \Cref{thm:main} under
Assumption~\ref{ass:strict-supermodularity}]
    Choose an extreme optimal solution \(\lambda\) to
    \eqref{eq:rco-lp}, and let \(\mathcal C=\supp(\lambda)\).
    By \Cref{prop:extreme-optimal-regular}, \(\mathcal C\) is
    \(A\)-regular. By \Cref{lem:a-regular-supports-lambda}, an $A$-regular collection supports a unique solution and its
    nonzero weights are \(1/(|\mathcal C|-1)\). Since every
    \(A\)-regular collection also supports a feasible fractional
    partition, maximizing over all such collections gives the
    expression in \Cref{thm:main}.
\end{proof}
    


It remains to prove \Cref{prop:extreme-optimal-regular}. Start with
an arbitrary extreme optimal solution
\(\lambda\in\Lambda(A)\) with support \(\mathcal C\).
By \Cref{lem:extreme-support}, \(\mathcal C\) is minimal among the
supports of elements of \(\Lambda(A)\). It therefore suffices to
find an \(A\)-regular subcollection
\(\mathcal C'\subseteq\mathcal C\). By
\Cref{lem:a-regular-supports-lambda}, \(\mathcal C'\) supports a
feasible fractional partition, so the minimality of \(\mathcal C\)
forces \(\mathcal C'=\mathcal C\). Thus \(\mathcal C\) itself is
\(A\)-regular.

The covering property in
\Cref{lem:optimal-incomparable-cover} is the main tool in constructing
\(\mathcal C'\). The four cases below differ only in the choice of
this subcollection.


    \subsection{\texorpdfstring{\(|A|=m\)}{|A|=m}}
    
    
    We begin with the no-helper case, for which the construction is
    the most straightforward. We need the notion of a \emph{minimal}
    set: a set \(C\in\mathcal C\) is minimal in \(\mathcal C\) if
    there is no \(C'\in\mathcal C\) such that \(C'\subsetneq C\).
    
    \begin{proof}[Proof of \Cref{prop:extreme-optimal-regular} for \(|A|=m\)]
        Let \(\lambda\) be any extreme optimal solution to
        \eqref{eq:rco-lp}, and let \(\mathcal C=\supp(\lambda)\).
        Let \(\mathcal C'\subseteq\mathcal C\) consist of all
        minimal sets of \(\mathcal C\). We must have $|\mathcal{C}'| \geq 2$. To see this,
        suppose by contradiction that $\mathcal{C}' = \{C_0\}$, $C_0 \neq \emptyset$. Then, $C_0 \subseteq C$ 
        for every $C \in \mathcal{C}$, contradicting \Cref{lem:support-of-feasible-sol}(ii).
    
        Distinct sets in \(\mathcal C'\) are incomparable. Hence
        the covering property in
        \Cref{lem:optimal-incomparable-cover} gives
        \begin{equation}\label{eq:cover}
            C\cup C'=[m]
            \qquad
            \text{for distinct }C,C'\in\mathcal C'.
        \end{equation}
    
        Fix \(x\in[m]\). By
        \Cref{lem:support-of-feasible-sol}(ii), some set in
        \(\mathcal C\) omits \(x\). Every set in a finite collection
        contains a minimal set of that collection, so some set in
        \(\mathcal C'\) also omits \(x\). On the other hand,
        \eqref{eq:cover} shows that two distinct sets in
        \(\mathcal C'\) cannot both omit \(x\). Thus every terminal
        is omitted from exactly one set in \(\mathcal C'\).
    
        Since \(A=[m]\), there are no helpers, and this proves that
        \(\mathcal C'\) is \(A\)-regular. 
        A strict
        inclusion \(\mathcal C'\subsetneq\mathcal C\) would
        contradict the extremality of \(\lambda\) by
        \Cref{lem:extreme-support}. Therefore,
        \(\mathcal C'=\mathcal C\), and the support of \(\lambda\)
        is \(A\)-regular.
    
    \end{proof}

    \subsection{Constructing $A$-regular subcollections}

        When \(|A|=m\), the minimal sets of the support of an optimal
        solution form a natural \(A\)-regular subcollection. In the presence
        of helpers, the same argument still shows that every target
        terminal is omitted from exactly one minimal set. A helper,
        however, may be omitted from several minimal sets and need not
        be matched with a target terminal. To address this difficulty,
        we consider sets in the support that omit a given target terminal.
        We first show that these sets form a nested collection and then
        use the maximum sets of such collections as the first step to constructing 
        the desired $A$-regular subcollection.
        
        A collection of sets is called \emph{nested} if every pair of
        its sets is comparable. Equivalently, its sets form a chain
        under inclusion. A finite nonempty nested collection has a
        unique minimum set and a unique maximum set.
        
        For a collection \(\mathcal C\subseteq 2^{[m]}\) and \(i\in[m]\),
        define
        \[
            \mathcal D_{\neg i}(\mathcal C)
            \coloneqq
            \{C\in\mathcal C:i\notin C\}.
        \]
        When \(\mathcal C\) is clear
        from the context, we simply write \(\mathcal D_{\neg i}\).
        
        \begin{lemma}\label{lem:mT_i}
            Under Assumption~\ref{ass:strict-supermodularity}, let
            \(\lambda\) be an optimal solution to \eqref{eq:rco-lp},
            and let \(\mathcal C=\supp(\lambda)\). For every \(i\in A\),
            the collection \(\mathcal D_{\neg i}\) is nonempty and
            nested.
        \end{lemma}
        
        \begin{proof}
            Fix \(i\in A\). The collection \(\mathcal D_{\neg i}\)
            is nonempty by \Cref{lem:support-of-feasible-sol}(ii).
            If two of its sets \(C,C'\) were incomparable, the
            covering property in \Cref{lem:optimal-incomparable-cover}
            would imply \(A\subseteq C\cup C'\), which is impossible
            because both sets omit \(i\in A\). Thus every pair of
            sets in \(\mathcal D_{\neg i}\) is comparable, proving
            that the collection is nested.
        \end{proof}
        
        For a support \(\mathcal C\) of an optimal solution and each \(i\in A\),
        let \(D_{\neg i}\) denote the unique maximum set of
        \(\mathcal D_{\neg i}\). The next
        lemma identifies exactly which terminals are omitted from
        these maximum sets.
        
        \begin{lemma}\label{lem:D-i-properties}
            Let \(\lambda\) be an optimal solution to \eqref{eq:rco-lp},
            let \(\mathcal C=\supp(\lambda)\), and let
            \(D_{\neg i}\), \(i\in A\), be defined as above.
            \begin{enumerate}[label=(\roman*)]
                \item For every \(i\in A\) and \(j\in[m]\),
                \[
                    i\text{ and }j\text{ are matched in }\mathcal C
                    \quad\Longleftrightarrow\quad
                    j\notin D_{\neg i}.
                \]
                \item For every \(i,j\in A\),
                \[
                    i\text{ and }j\text{ are matched in }\mathcal C
                    \quad\Longleftrightarrow\quad
                    D_{\neg i}=D_{\neg j}.
                \]
                \item Every helper that is not matched in \(\mathcal C\)
                with any target terminal belongs to every
                \(D_{\neg i}\), \(i\in A\).
            \end{enumerate}
        \end{lemma}
        
        \begin{proof}
            To prove (i), fix \(i\in A\) and \(j\in[m]\). If \(i\) and \(j\)
are matched in \(\mathcal C\), then they are contained in exactly
the same members of \(\mathcal C\). Hence,
\(i\notin D_{\neg i}\) implies \(j\notin D_{\neg i}\).
        
            Conversely, suppose that \(j\notin D_{\neg i}\).
            Every \(C\in\mathcal D_{\neg i}\) satisfies
            \(C\subseteq D_{\neg i}\), so every such \(C\) also
            omits \(j\). Therefore,
            \[
                \mathcal D_{\neg i}
                \subseteq
                \mathcal D_{\neg j}.
            \]
            The balance conditions \Cref{eq:balance} give
            \[
                \sum_{C\in\mathcal D_{\neg i}}\lambda_C
                =
                \sum_{C\in\mathcal C}\lambda_C-1
                =
                \sum_{C\in\mathcal D_{\neg j}}\lambda_C.
            \]
            Since every set in \(\mathcal C\) has strictly positive
            weight, the inclusion must be an equality. Thus the
            sets omitting \(i\) are exactly those omitting \(j\),
            which means that \(i\) and \(j\) are matched in
            \(\mathcal C\). This proves (i).
        
            To prove (ii), let \(i,j\in A\). If \(i\) and \(j\)
            are matched, then
            \(\mathcal D_{\neg i}=\mathcal D_{\neg j}\), so their
            maximum sets are equal. Conversely, if
            \(D_{\neg i}=D_{\neg j}\), then \(j\notin D_{\neg i}\),
            and (i) shows that \(i\) and \(j\) are matched.
        
            Finally, (iii) follows from (i): if a helper \(j\)
            were omitted from some \(D_{\neg i}\), \(i\in A\),
            then it would be matched with the target terminal \(i\).
        \end{proof}

        \begin{corollary}\label{cor:maximum-sets}
            In the collection of distinct maximum sets
            \[
                \mathcal D_A
                \coloneqq
                \{D_{\neg i}:i\in A\}
                \subseteq\mathcal C,
            \]
            every target terminal \(i \in A\) is omitted from exactly one set in \(\mathcal D_A\), namely \(D_{\neg i}\). Further, $|\mathcal{D}_A| \geq 2$.
        \end{corollary}
        \begin{proof}
            If $i \notin D_{\neg j}$ for some $j \in A$, $ j\neq i$, then
            Lemma~\ref{lem:D-i-properties}(i) implies that $i$ and $j$
            are matched in $\mathcal C$. Hence, by
            Lemma~\ref{lem:D-i-properties}(ii),
            $D_{\neg i} = D_{\neg j}$. 
            
            We prove $|\mathcal{D}_A| \geq 2$ by contradiction: suppose that $\mathcal{D}_A = \{D\}$, and let $i \neq j \in A$. Because $D_{\neg i} = D_{\neg j}$, \Cref{lem:D-i-properties}(ii) implies $i$ and $j$ are matched in $\mathcal C$. Every element of $\mathcal{C}$ omits at least one element of $A$, and so must omit \emph{every} element of $A$. This contradicts the balance condition \Cref{eq:balance} for any $i \in A$.
        \end{proof}

        If each helper terminal was matched in $\mathcal{C}$ with some terminal in $A$, \Cref{cor:maximum-sets} would imply  that \(\mathcal D_A\) is \(A\)-regular and our construction would be complete. However, if there is a helper terminal that is not matched in $\mathcal{C}$ with any target terminal, that helper terminal belongs to every set in $\mathcal{D}_A$ by \Cref{lem:D-i-properties}(iii), and so $\mathcal{D}_A$ is not $A$-regular. The following lemmas help to address this issue, and will be used in our proofs of the remaining cases.
        

        
        \begin{lemma}\label{lem:support-of-optimal-intersects}
            Under Assumption~\ref{ass:strict-supermodularity}, if
            \(\lambda\) is an optimal solution to \eqref{eq:rco-lp},
            then every \(C\in\supp(\lambda)\) satisfies
            \(C\cap A\neq\emptyset\).
        \end{lemma}
        
        \begin{proof}
            Let \(\mathcal C=\supp(\lambda)\), and suppose that some
            \(C\in\mathcal C\) satisfies \(C\cap A=\emptyset\).
            Choose a minimal set \(C_0\in\mathcal C\) contained in
            \(C\). Then \(C_0\cap A=\emptyset\).
        
            If some \(C'\in\mathcal C\) were incomparable with
            \(C_0\), the covering property in
            \Cref{lem:optimal-incomparable-cover} would give
            \(A\subseteq C_0\cup C'\).
            Since \(C_0\cap A=\emptyset\),
            this would imply \(A\subseteq C'\), contrary to
            \(C'\in\mathcal B(A)\). 
            Thus every set in
            \(\mathcal C\) is comparable with \(C_0\).
        
            By the minimality of \(C_0\), it follows that
            \(C_0\subseteq C'\) for every \(C'\in\mathcal C\).
            Since \(C_0\neq\emptyset\), any element of \(C_0\)
            belongs to every set in \(\mathcal C\), contradicting
            \Cref{lem:support-of-feasible-sol}(ii).
        \end{proof}

        
        \begin{lemma}\label{lem:common-helper-omission}
            Let \(\lambda\) be an optimal solution to \eqref{eq:rco-lp},
            and let \(\mathcal C=\supp(\lambda)\). Consider the helper terminals that 
            are not matched in $\mathcal{C}$ with any element of $A$. If there exists
            a set \(C_0\in\mathcal C\) that omits every such helper terminal, the subcollection
            \[
                \mathcal C'
                \coloneqq
                \{C_0\}
                \cup
                \{D_{\neg i}:i\in A\cap C_0\} \subseteq \mathcal C
            \]
            is \(A\)-regular.
        \end{lemma}

        \begin{remark}
    Such a $C_0$ may not always exist, but in the specific cases
    we consider below, we shall show its existence.
    If $C_0$ exists, the set $A \cap C_0$ is nonempty by
    \Cref{lem:support-of-optimal-intersects}, and $C_0$ is
    distinct from every selected $D_{\neg i}$ because
    $i\in C_0$ but $i\notin D_{\neg i}$ for every
    $i\in A\cap C_0$. Thus, $|\mathcal C'|\ge 2$.
\end{remark}
        
        \begin{proof}
            We first verify condition~(i) of \(A\)-regularity, that every
            target terminal is omitted from exactly one set in \(\mathcal C'\).
            Fix \(i\in A\) and consider two cases.
        
            \textbf{Case 1:} \(i\in A\setminus C_0\).
            For every \(j\in A\cap C_0\), the set \(C_0\) contains
            \(j\) but omits \(i\), so they are not matched in
            \(\mathcal C\). By \Cref{lem:D-i-properties}(i),
            \(i\in D_{\neg j}\). Thus \(i\) is omitted from exactly
            one set in \(\mathcal C'\), namely \(C_0\). 
        

            \textbf{Case 2:} \(i\in A\cap C_0\).
            Then \(i\in C_0\) and \(i\notin D_{\neg i}\).
            For any \(j\in A\cap C_0\) such that
            \(D_{\neg j}\neq D_{\neg i}\), the terminals \(i\) and \(j\)
            are not matched by \Cref{lem:D-i-properties}(ii), and so
            \(i\in D_{\neg j}\) by \Cref{lem:D-i-properties}(i).
            Hence \(i\) is omitted from exactly one distinct set
            in \(\mathcal C'\), namely \(D_{\neg i}\).
            This verifies condition~(i) of $A$-regularity.
        
            To verify condition~(ii) of $A$-regularity, observe that any helper already
            matched with a target in \(\mathcal C\) is matched to the same target 
            in \(\mathcal C'\). Every other helper
            is omitted from \(C_0\) by hypothesis and belongs to
            every selected maximum set by
            \Cref{lem:D-i-properties}(iii). Now consider any target terminal $i \in A \setminus C_0$, which is nonempty because
            \(C_0\in\mathcal B(A)\). The terminal $i$ $\notin C_0$ by definition and, for any $j \in A \cap C_0$, $i$ and $j$ are not matched in $\mathcal{C}$, which implies $i \in D_{\neg j}$ by \Cref{lem:D-i-properties}(i). Therefore, every such helper is matched
            in \(\mathcal C'\) with every element of
            \(A\setminus C_0\). 
            
            Thus, \(\mathcal C'\) is \(A\)-regular.
        \end{proof}

        \subsection{\texorpdfstring{\(|A|=m-1\)}{|A|=m-1}}
        
        \begin{proof}[Proof of \Cref{prop:extreme-optimal-regular} for \(|A|=m-1\)]
            Without loss of generality, let \(A=[m-1]\).
            Let \(\lambda\) be an extreme optimal solution to
            \eqref{eq:rco-lp}, and let
            \(\mathcal C=\supp(\lambda)\). We consider two cases.

            \textbf{Case 1:} Suppose the helper \(m\) is matched in \(\mathcal C\) with a 
            target terminal $i \in A$, which implies that it will be matched with $i$ in any 
            subcollection of $\mathcal{C}$. Consider the subcollection $\mathcal{C}' = \mathcal{D}_A \subseteq \mathcal{C}$.
            By \Cref{cor:maximum-sets}, every $j \in A$ is omitted from exactly one set in $\mathcal{C}'$. 
            Thus, $\mathcal{C}'$ is $A$-regular.

        
            \textbf{Case 2:} Suppose the helper \(m\) is not matched in \(\mathcal C\) with \emph{any} 
            terminal in $A$. By \Cref{lem:support-of-feasible-sol}(ii), there exists 
            a set \(C_0\in\mathcal C\) with \(m\notin C_0\).
            Since \(m\) is the only helper, \(C_0\) satisfies the
            hypothesis of \Cref{lem:common-helper-omission}.
            Consequently,
            \[
                \mathcal C'
                =
                \{C_0\}
                \cup
                \{D_{\neg i}:i\in A\cap C_0\}
            \]
            is \(A\)-regular.
        
            In either case, we have found an \(A\)-regular subcollection
            \(\mathcal C'\subseteq\mathcal C\). 
            A strict inclusion
            \(\mathcal C'\subsetneq\mathcal C\) would contradict
            the extremality of \(\lambda\) by
            \Cref{lem:extreme-support}. Therefore,
            \(\mathcal C'=\mathcal C\), so the support of \(\lambda\)
            is \(A\)-regular.
        \end{proof}
        
        \subsection{\texorpdfstring{\(|A|=m-2\)}{|A|=m-2}}
        
        \begin{proof}[Proof of \Cref{prop:extreme-optimal-regular} for \(|A|=m-2\)]
            Without loss of generality, let
            \[
                A=[m-2],
                \qquad
                A^c=\{m-1,m\}.
            \]
            Let \(\lambda\) be an extreme optimal solution to
            \eqref{eq:rco-lp}, and let
            \(\mathcal C=\supp(\lambda)\). As in the case $|A| = m-1$, we will construct
            an $A$-regular subcollection of $\mathcal{C}$.

            If both helper terminals are matched with (not necessarily the same) 
            target terminals, by \Cref{cor:maximum-sets}, $\mathcal{C}' = \mathcal{D}_A$
            is an $A$-regular subcollection.
        
        
            If exactly one helper terminal is not matched with any target terminal,
            we may assume that this helper is \(m\). As in Case 2 above when $|A|=m-1$, we can choose
            some \(C_0\in\mathcal C\) with \(m\notin C_0\). Such a $C_0$ exists by
            \Cref{lem:support-of-feasible-sol}(ii). Since the
            other helper \(m-1\) is matched with a target terminal,
            \(C_0\) omits every helper not matched with a target. Thus, by \Cref{lem:common-helper-omission},
            \[
                \mathcal C'
                =
                \{C_0\}
                \cup
                \{D_{\neg i}:i\in A\cap C_0\} \subseteq \mathcal{C}
            \]
            is an \(A\)-regular subcollection of \(\mathcal C\).
        
            It remains to consider the case in which neither helper
            is matched in \(\mathcal C\) with an element of \(A\).
            If some \(C_0\in\mathcal C\) omits both \(m-1\) and \(m\),
            then \Cref{lem:common-helper-omission} again gives an
            \(A\)-regular subcollection. We may therefore assume that
            no set in \(\mathcal C\) omits both helpers. 
            
        
            By \Cref{lem:support-of-feasible-sol}(ii), choose
            \(C_{\neg(m-1)},C_{\neg m}\in\mathcal C\) such that
            \[
                m-1\notin C_{\neg(m-1)}
                \ \text{and}\
                m\notin C_{\neg m}.
            \]
            Since we have assumed that no set in $\mathcal{C}$ omits both helpers, we must have
            \begin{equation}\label{eq:cross-membership}
                m\in C_{\neg(m-1)}
                \ \text{and}\
                m-1\in C_{\neg m}.
            \end{equation}
            Hence $C_{\neg m}, C_{\neg (m-1)}$ are incomparable and the covering
            property in \Cref{lem:optimal-incomparable-cover} gives
            \[
                A\subseteq C_{\neg(m-1)}\cup C_{\neg m},
            \]
            or, equivalently,
            \begin{equation}\label{eq:disjoint-omissions}
                (A\setminus C_{\neg(m-1)})
                \cap
                (A\setminus C_{\neg m})
                =
                \emptyset.
            \end{equation}
        
            Define
            \begin{equation}\label{eq:mminus2-two-sets}
                \mathcal C'
                \coloneqq
                \{C_{\neg(m-1)},C_{\neg m}\}
                \cup
                \{D_{\neg i}:
                    i\in A\cap C_{\neg(m-1)}\cap C_{\neg m}\}.
            \end{equation}
            We claim that \(\mathcal C'\) is \(A\)-regular. The intersection \(A\cap C_{\neg(m-1)}\cap C_{\neg m}\) may be empty, in which case \(\mathcal C'\) consists only of the two sets \(C_{\neg(m-1)}\) and \(C_{\neg m}\); the argument below applies unchanged.
        
            Let $ i \in A$. By \Cref{eq:disjoint-omissions}, $A$ is the disjoint union of 
            \begin{align*}
                A\setminus C_{\neg(m-1)}, \quad A\setminus C_{\neg m}, \quad \text{and} \quad A\cap C_{\neg(m-1)}\cap C_{\neg m}.
            \end{align*}
        
            If \(i\in A\setminus C_{\neg(m-1)}\), 
            then \(i\in C_{\neg m}\). Further, for every
            \(j\in A\cap C_{\neg(m-1)}\cap C_{\neg m}\), the set
            \(C_{\neg(m-1)}\) contains \(j\) but omits \(i\).
            Thus \(i\) and \(j\) are not matched in \(\mathcal C\),
            and \Cref{lem:D-i-properties}(i) gives
            \(i\in D_{\neg j}\). Hence \(i\) belongs to every
            selected maximum set and is omitted only from
            \(C_{\neg(m-1)}\). The case \(i\in A\setminus C_{\neg m}\) 
            can be handled similarly.
        
            The case 
            \(i\in A\cap C_{\neg(m-1)}\cap C_{\neg m}\) remains. Here,
            \(i\) belongs to both \(C_{\neg(m-1)}\) and
            \(C_{\neg m}\), while \(D_{\neg i}\) is among the
            selected maximum sets and omits \(i\). If another
            selected maximum set \(D_{\neg j}\) is distinct from
            \(D_{\neg i}\), then \(i\) and \(j\) are not matched
            by \Cref{lem:D-i-properties}(ii), so
            \(i\in D_{\neg j}\) by \Cref{lem:D-i-properties}(i).
            Thus \(i\) is omitted from exactly one distinct set
            in \(\mathcal C'\), namely \(D_{\neg i}\).
        
            In all three cases, the target terminal \(i\) is omitted
            from exactly one set in \(\mathcal C'\). This verifies
            condition~(i) of $A$-regularity.
        
            To verify condition~(ii), recall that neither helper is
            matched with a target in \(\mathcal C\). Thus both
            helpers belong to every \(D_{\neg i}\) by
            \Cref{lem:D-i-properties}(iii). Together with
            \eqref{eq:cross-membership}, this shows that \(m-1\)
            is omitted only from \(C_{\neg(m-1)}\), while \(m\)
            is omitted only from \(C_{\neg m}\). Therefore,
            \(m-1\) is matched in \(\mathcal C'\) with every
            element of \(A\setminus C_{\neg(m-1)}\), and \(m\)
            is matched with every element of \(A\setminus C_{\neg m}\).
            Both \(A\setminus C_{\neg(m-1)}\), \(A\setminus C_{\neg m}\) are nonempty because 
             \(C_{\neg m}, C_{\neg m-1} \in \mathcal B(A)\). This verifies
            condition~(ii), so \(\mathcal C'\) is \(A\)-regular.
        
            In all cases, we have constructed an \(A\)-regular
            subcollection \(\mathcal C'\subseteq\mathcal C\).
            By \Cref{lem:a-regular-supports-lambda}, this subcollection
            supports a feasible fractional partition. A strict
            inclusion \(\mathcal C'\subsetneq\mathcal C\) would
            contradict the extremality of \(\lambda\) by
            \Cref{lem:extreme-support}. Therefore,
            \(\mathcal C'=\mathcal C\), so the support of \(\lambda\)
            is \(A\)-regular, proving this case of \Cref{prop:extreme-optimal-regular}.
        \end{proof}
        
        \subsection{\texorpdfstring{\(|A|=2\)}{|A|=2}}
        
        
        \begin{proof}[Proof of \Cref{prop:extreme-optimal-regular} for \(|A|=2\)]
            Without loss of generality, let \(A=\{1,2\}\).
            Let \(\lambda\) be an extreme optimal solution to
            \eqref{eq:rco-lp}, and let
            \(\mathcal C=\supp(\lambda)\).
        
            By \Cref{lem:support-of-feasible-sol}(i), no set in
            \(\mathcal C\) contains both \(1\) and \(2\), while
            \Cref{lem:support-of-optimal-intersects} shows that every
            set contains at least one of them.
            Thus every set contains exactly one target terminal,
            and \(\mathcal C\) is a disjoint union of
            \[
                \mathcal C_1
                \coloneqq
                \{C\in\mathcal C:1\in C\}
                =
                \mathcal D_{\neg2},
                \qquad
                \mathcal C_2
                \coloneqq
                \{C\in\mathcal C:2\in C\}
                =
                \mathcal D_{\neg1}.
            \]
            By \Cref{lem:mT_i}, both collections are nonempty and nested.
        
            Let \(C_1\) be the maximum set of \(\mathcal C_1\),
            and let \(C_2\) be the minimum set of \(\mathcal C_2\).
            We show that \(C_1\) and \(C_2\) partition \([m]\).
        
            We first show that $C_1 \cap C_2 = \emptyset$. Suppose not, and let \(x\in C_1\cap C_2\).
            Since \(C_2\) is the minimum set of \(\mathcal C_2\),
            \(x\) belongs to every set in \(\mathcal C_2\).
            In addition, \(x\in C_1\), where
            \(C_1\notin\mathcal C_2\). Therefore,
            \[
                \sum_{\substack{C\in\mathcal C\\x\in C}}\lambda_C
                \geq
                \sum_{C\in\mathcal C_2}\lambda_C+\lambda_{C_1}
                =
                1+\lambda_{C_1}
                >1,
            \]
            where the equality uses the balance condition \Cref{eq:balance} for
            terminal \(2\). This contradicts the balance condition
            for \(x\). Hence \(C_1\cap C_2=\emptyset\).
        
            Next, suppose that \(x\notin C_1\cup C_2\).
            Since \(C_1\) is the maximum set of \(\mathcal C_1\),
            \(x\) belongs to no set in \(\mathcal C_1\).
            Moreover, \(x\notin C_2\), so
            \[
                \sum_{\substack{C\in\mathcal C\\x\in C}}\lambda_C
                \leq
                \sum_{C\in\mathcal C_2\setminus\{C_2\}}\lambda_C
                =
                1-\lambda_{C_2}
                <1,
            \]
            again contradicting the balance condition \Cref{eq:balance} for \(x\).
            Therefore, \(C_1\cup C_2=[m]\).
        
            Thus \(C_1,C_2\) form a partition of \([m]\), with
            \(1\in C_1\) and \(2\in C_2\). Since each set is the
            complement of the other, the collection
            \(\mathcal C'=\{C_1,C_2\}\) is \(A\)-regular by
            \Cref{lem:partition-A-regular}.
        
            By \Cref{lem:a-regular-supports-lambda},
            \(\mathcal C'\) supports a feasible fractional partition.
            Since \(\mathcal C'\subseteq\mathcal C\), the extremality
            of \(\lambda\) and \Cref{lem:extreme-support} imply
            \(\mathcal C'=\mathcal C\). Hence the support of
            \(\lambda\) is \(A\)-regular.
        \end{proof}

    \section{Dropping the assumption of strict supermodularity}
    \label{sec:non-strict-supermodularity}
    
        In \Cref{sec:proofs}, we proved \Cref{thm:main} under
        Assumption~\ref{ass:strict-supermodularity}. We now remove this
        assumption by augmenting the source with auxiliary random variables
        of arbitrarily small total entropy. The augmented source
        \(X_{[m]}^\epsilon\) has a strictly supermodular conditional-entropy
        function. As \(\epsilon\downarrow0\), its conditional entropies and
        minimum CO rate converge to those of the original source, allowing
        us to recover the desired conclusion.
        
        \begin{lemma}\label{lem:not-strict-supermodular}
            Let \(\epsilon>0\). For each unordered pair
            \(\{i,j\}\subseteq[m]\), $i \neq j$, introduce a finite-valued rv
            \(\zeta_{\{i,j\}}\) of entropy \(\epsilon/\binom{m}{2}\).
            These rvs are mutually independent and independent of
            \(X_{[m]}\). Set
            \[
                \zeta_i=(\zeta_{\{i,j\}}:j\in[m]\setminus\{i\}),
                \qquad
                X_i^\epsilon=(X_i,\zeta_i),
                \qquad i\in[m].
            \]
            Then the function 
            \( B \longmapsto H(X_B^\epsilon\,|\,X_{B^c}^\epsilon)\)
            is strictly supermodular, and
            \[
                R_{\mathrm{CO}}(A;X_{[m]})
                \leq
                R_{\mathrm{CO}}(A;X_{[m]}^\epsilon)
                \leq
                R_{\mathrm{CO}}(A;X_{[m]})+\epsilon.
            \]
        \end{lemma}
        
        \begin{proof}
            By independence, for every \(B\subseteq[m]\),
            \[
                H(X_B^\epsilon\,|\,X_{B^c}^\epsilon)
                =
                H(X_B\,|\,X_{B^c})
                +
                H(\zeta_B\,|\,\zeta_{B^c}).
            \]
            Conditioning on \(\zeta_{B^c}\) reveals every auxiliary variable
            associated with a pair of terminals having at least one endpoint in
            \(B^c\), leaving only those associated with pairs contained in \(B\).
            By independence,
            \begin{equation}\label{eq:auxiliary-conditional-entropy}
                H(\zeta_B\,|\,\zeta_{B^c})
                =
                \sum_{\{i,j\}\subseteq B}H(\zeta_{\{i,j\}})
                =
                \frac{\epsilon}{\binom{m}{2}}\binom{|B|}{2}.
            \end{equation}
            Consequently, for incomparable \(B,B'\subseteq[m]\),
            \begin{align*}
                H(\zeta_{B\cup B'}\,|\,\zeta_{(B\cup B')^c})
                +H(\zeta_{B\cap B'}\,|\,\zeta_{(B\cap B')^c})
                -H(\zeta_B\,|\,\zeta_{B^c})
                -H(\zeta_{B'}\,|\,\zeta_{(B')^c}) =
                \frac{\epsilon}{\binom{m}{2}}
                |B\setminus B'|\,|B'\setminus B|
                >0.
            \end{align*}
            Thus \(B\mapsto H(\zeta_B\,|\,\zeta_{B^c})\) is strictly
            supermodular. Since \(B\mapsto H(X_B\,|\,X_{B^c})\) is
            supermodular, the decomposition above shows that
            \(B\mapsto H(X_B^\epsilon\,|\,X_{B^c}^\epsilon)\) is
            strictly supermodular.
        
            For the lower bound, any protocol achieving omniscience for
            \(X_{[m]}^\epsilon\) also achieves omniscience for \(X_{[m]}\),
            which directly gives the desired lower bound.
        
            For the upper bound, omniscience can be attained separately
            for the original source and the auxiliary variables.
            Running the two protocols consecutively gives
            \[
                R_{\mathrm{CO}}(A;X_{[m]}^\epsilon)
                \leq
                R_{\mathrm{CO}}(A;X_{[m]})
                +
                R_{\mathrm{CO}}(A;\zeta_{[m]})
                \leq
                R_{\mathrm{CO}}(A;X_{[m]})+\epsilon.
            \]
            The last inequality follows by source coding each pair-variable
            sequence at one of its two terminals and broadcasting it, which implies
            $R_{\mathrm{CO}}(A;\zeta_{[m]}) \leq H(\zeta_{[m]}) = \epsilon$.
        \end{proof}
        
        For \(|A|\in\{2,m-2,m-1,m\}\), the result of \Cref{sec:proofs}
        applies to \(X_{[m]}^\epsilon\). By
        \eqref{eq:auxiliary-conditional-entropy}, each conditional entropy
        converges to its counterpart for \(X_{[m]}\) as $\epsilon$ approaches $0$. Since the family of
        \(A\)-regular collections is finite and does not depend on
        \(\epsilon\), the limit can be interchanged with the maximum.
        Together with \Cref{lem:not-strict-supermodular}, this gives
        \begin{align*}
            R_{\mathrm{CO}}(A;X_{[m]})
            &=
            \lim_{\epsilon\downarrow0}
            R_{\mathrm{CO}}(A;X_{[m]}^\epsilon)\\
            &=
            \lim_{\epsilon\downarrow0}
            \max_{\substack{
                \mathcal C\subseteq\mathcal B(A)\\
                \mathcal C\text{ is }A\text{-regular}
            }}
            \frac{1}{|\mathcal C|-1}
            \sum_{C\in\mathcal C}
            H(X_C^\epsilon\,|\,X_{C^c}^\epsilon)\\
            &=
            \max_{\substack{
                \mathcal C\subseteq\mathcal B(A)\\
                \mathcal C\text{ is }A\text{-regular}
            }}
            \frac{1}{|\mathcal C|-1}
            \sum_{C\in\mathcal C}H(X_C\,|\,X_{C^c}).
        \end{align*}
        Any maximizing collection supports an optimal fractional partition
        by \Cref{lem:a-regular-supports-lambda}. This proves \Cref{thm:main}
        and, because SK capacity and the minimum CO rate are related as in \eqref{eq:si-rco-rel} and \eqref{eq:partition-identity},
        the positive cases of \Cref{thm:SK-cap-main}.
        
        \begin{remark}\label{rem:general-supermodular-objective}
        The proof above uses the communication interpretation when extending
        the result beyond strict supermodularity. However, this interpretation
        is not essential. Observe that the arguments in \Cref{sec:proofs}
        apply to any strictly supermodular set function
        \(f:2^{[m]}\to\mathbb R\) satisfying \(f(\emptyset)=0\).
        For \(|A|\in\{2,m-2,m-1,m\}\), it shows that the linear program
        \[
            \max_{\lambda\in\Lambda(A)}
            \sum_{B\in\mathcal B(A)}\lambda_B f(B)
        \]
        has an optimal solution with an \(A\)-regular support.
        
        In particular, for any \(\epsilon>0\), consider
        \[
            f(B)
            =
            H(X_B\,|\,X_{B^c})
            +
            \frac{\epsilon}{\binom{m}{2}}
            \binom{|B|}{2},
            \qquad B\subseteq[m].
        \]
        As shown above, \(f\) is strictly supermodular and satisfies
        \(f(\emptyset)=0\). Thus \Cref{sec:proofs} establishes the result
        for the corresponding LP for every \(\epsilon>0\).
        Applying the same limiting argument as \(\epsilon\downarrow0\)
        then gives the result for the original conditional-entropy
        objective, without using its communication interpretation.
        \end{remark}
    
    \section{Counterexamples for \texorpdfstring{\(3\leq |A|\leq m-3\)}{3 <= |A| <= m-3}} \label{sec:counterexample}
        For \(m=6\) and three target terminals, Chan
\cite{chan2008tightness} gave the following example showing that
the lower bound on the minimum CO rate in the inequality \Cref{eq:partition-rco-bound} can be strict.
Exploiting the relationship to SK capacity \Cref{eq:si-rco-rel}, this further implies that the partition
upper bound in \Cref{eq:upper-bound} can be strict.
        
        \begin{proposition}[\cite{chan2008tightness}]\label{prop:chan}
            Let \(m=6\), let \(\tilde A=\{1,2,3\}\), and let
            \(Y_1,Y_2,Y_3,Y_4\) be independently and uniformly distributed bits.
            Define
            \[
                \begin{aligned}
                    \tilde X_1&\coloneqq Y_1\oplus Y_3,
                    &\tilde X_4&\coloneqq Y_2\oplus Y_3,\\
                    \tilde X_2&\coloneqq Y_1\oplus Y_4,
                    &\tilde X_5&\coloneqq Y_2\oplus Y_4,\\
                    \tilde X_3&\coloneqq Y_3\oplus Y_4,
                    &\tilde X_6&\coloneqq Y_1\oplus Y_2.
                \end{aligned}
            \]
            Then \(H(\tilde X_{[6]})=3\), and\footnote{In \cite{chan2008tightness} it is incorrectly claimed that for the rvs $\tilde{X}_{[6]}$, $H(\tilde{X}_{B} \,|\, \tilde{X}_{B^c})$ depends only on $|B|$. While this is not true, the conclusion \Cref{eq:strict-inequality} still holds.}
            \begin{equation}\label{eq:strict-inequality}
                R_{\mathrm{CO}}(\tilde A;\tilde X_{[6]})
                \geq
                \frac94
                >
                \max_{\substack{
                    \mathcal C\subseteq\mathcal B(\tilde A)\\
                    \mathcal C\text{ is }\tilde A\text{-regular}
                }}
                \frac{1}{|\mathcal C|-1}
                \sum_{C\in\mathcal C}
                H(\tilde X_C\,|\,\tilde X_{C^c})
                =
                2.
            \end{equation}
        \end{proposition}
        
        We extend this example to arbitrary $m$ and $A$ not covered by the positive cases of \Cref{thm:main} by duplicating observations across suitable
        groups of terminals. The construction preserves the optimized
        partition bound while not decreasing the minimum CO rate.

        \begin{proposition}\label{prop:counterexample-extension}
    For every \(A\subseteq[m]\) with
    \(3\leq |A|\leq m-3\), there exists a finite-alphabet
    DMMS for which the minimum CO rate lower bound
    \eqref{eq:partition-rco-bound} is strict.
\end{proposition}
        
        \begin{proof}
            Partition \(A\) into three nonempty sets \(A_1,A_2,A_3\),
            and \(A^c\) into three nonempty sets \(A_4,A_5,A_6\).
            Using the source in \Cref{prop:chan}, define
            \[
                X_j\coloneqq\tilde X_i
                \qquad\text{for }j\in A_i,\quad i\in[6].
            \]
            Thus all terminals in \(A_i\) observe the same rv $\tilde X_i$, and
            \(H(X_{[m]})=3\).
        
            Let \((R_1,\ldots,R_m)\) be any feasible rate vector in the LP 
        \eqref{eq:rco-primal} for \(X_{[m]}\). For the six-terminal
        source, assign to terminal \(i\) the total rate of the terminals
        in \(A_i\), namely
        \[
            \tilde R_i\coloneqq\sum_{j\in A_i}R_j,
            \qquad i\in[6].
        \]
        These rates satisfy the Slepian--Wolf constraints for
        \(\tilde X_{[6]}\): each constraint indexed by
        \(B\in\mathcal B(\tilde A)\) follows from the constraint
        indexed by \(\bigcup_{i\in B}A_i\) for \(X_{[m]}\), since
        repeating observations does not change the corresponding
        conditional entropies. The total rate is unchanged.
        
        Minimizing over feasible rate vectors for the expanded source
        therefore gives
        \begin{equation}\label{eq:extended-rco-lower-bound}
            R_{\mathrm{CO}}(A;X_{[m]})
            \geq
            R_{\mathrm{CO}}(\tilde A;\tilde X_{[6]})
            \geq
            \frac94,
        \end{equation}
        where the last inequality follows from \Cref{prop:chan}.
        
            We next show that the optimized partition lower bound
            remains exactly \(2\). A partition may split a class
            \(A_i\), so it is not enough to consider only partitions
            obtained from the six-terminal source.
        
            Let \(\pi\in\Pi(A)\) have \(k\) blocks. Every block
            contains a target terminal and therefore has entropy
            at least one. A block containing a helper has entropy
            at least two, since the observations of any target
            terminal and any helper are independent uniform bits.
            Either at least two blocks contain helpers, or all
            helpers lie in a single block, which has entropy three
            because it contains the three independent observations
            \(\tilde X_4,\tilde X_5,\tilde X_6\). In either case,
            \[
                \sum_{D\in\pi}H(X_D)\geq k+2,
            \]
            and hence
            \[
                \frac{1}{k-1}
                \sum_{D\in\pi}H(X_{D^c}\,|\,X_D)
                =
                \frac{kH(X_{[m]})-\sum_{D\in\pi}H(X_D)}{k-1} \leq \frac{3k-(k+2)}{k-1} = 2
            \]
            Equality is attained by the admissible partition
            \[
                \pi^\star
                =
                \{A_1,A_2,A_3\cup A_4\cup A_5\cup A_6\},
            \]
            whose block entropies are \(1,1,3\). Therefore, by
            \Cref{lem:partition-A-regular},
            \[
                \max_{\substack{
                    \mathcal C\subseteq\mathcal B(A)\\
                    \mathcal C\text{ is }A\text{-regular}
                }}
                \frac{1}{|\mathcal C|-1}
                \sum_{C\in\mathcal C}H(X_C\,|\,X_{C^c})
                =
                2.
            \]
            Combining this equality with \eqref{eq:extended-rco-lower-bound}
        shows that the partition lower bound equals \(2\), whereas
        the minimum CO rate is at least \(9/4\). Hence the lower
        bound is strict.
        \end{proof}
        
        This establishes the remaining cases of \Cref{thm:SK-cap-main}
        and completes the characterization.

    \section{Secret key capacity for Markov random fields with two target terminals}\label{sec:structural-A-2}
        As noted in the introduction, it was shown in \cite{csiszar2004secrecy} that the partition upper bound is tight when the DMMS is a Markov chain or, more generally, a tree-structured Markov random field and $|A| = m$. Further, the optimal partition is the one obtained by cutting the edge $(i^*,j^*)$ in the tree with the minimum mutual information, that is, 
        \begin{align}
            C_\mathrm{SK}([m]) = I(X_{i^*} \wedge X_{j^*}) = \min_{(i,j) \text{ is an edge}} I(X_i \wedge X_j). \label{eq:sk-trees}
        \end{align} 
        A different proof of this result is given in \cite{bhattacharya-narayan-mct}, which also exploits \Cref{eq:sk-trees} to propose an algorithm to estimate the SK capacity when the structure of the tree is known but the underlying joint distribution is unknown, reducing the search space for the optimal partition in \Cref{eq:sk-capacity} from the exponentially large (in $m$) space of all allowed partitions to the linear (in $m$) space of partitions obtained by cutting a tree-edge. Analogous results were also obtained for nontree Markov random fields in \cite{bhattacharya2023cliqueylon,pathegama2024markov}. Thus, structural insights into the optimization problem \Cref{eq:sk-capacity} enable more efficient estimation of the SK capacity when $P_{X_{[m]}}$ is unknown. 
        
        In this section, we investigate the optimization in \Cref{eq:sk-capacity} when the underlying DMMS is a Markov random field and $|A|=2$. One such property is already known: \cite[Eq. (36)]{csiszar2004secrecy} showed that for any $A$ and a tree-structured Markov random field, $C_\mathrm{SK}(A)$ is given by a similar expression as \Cref{eq:sk-trees}, but only considering edges that are on the minimal tree connecting every vertex in $A$. 

     
        We begin by defining a Markov random field, which associates the rvs $X_{[m]}$ with a suitable underlying graph $\mathcal G$ that encodes conditional independencies of $P_{X_{[m]}}$ using \emph{separations} in $\mathcal G$.
    
        \begin{definition}[Markov random field \cite{koller-friedman-pgm}]
            Let \(\mathcal G=([m],\mathcal E)\) be an undirected graph.
        For disjoint sets \(A, A', B\subseteq[m]\), with
        \(A,A'\neq\emptyset\), we say that \(B\) \emph{separates}
        \(A\) and \(A'\) if every path from a vertex in \(A\)
        to a vertex in \(A'\) intersects \(B\). 
    
        Assigning rv \(X_i\) to vertex \(i\), the pmf
        \(P_{X_{[m]}}\) satisfies the \emph{global Markov property}
        with respect to \(\mathcal G\) if the Markov chain
        \[
            X_{A}\mc X_B\mc X_{A'}
        \]
        holds whenever \(B\) separates \(A\) and \(A'\).
        In this case, \(X_{[m]}\) is called a \emph{Markov random
        field} with respect to \(\mathcal G\).
        \end{definition}
        Next, we restate \Cref{thm:SK-cap-main} for the case $|A|=2$ as follows: the secret key capacity when $|A| = 2$ is given by the minimum mutual information $I(X_B \wedge X_{B^c})$ across any bipartition $(B, B^c)$ of $[m]$ that separates the two terminals in $A$. The following is our main result in this section. It shows that for Markov random fields, the optimization in \Cref{eq:sk-capacity} for $|A| = 2$ can be restricted to only those bipartitions $\pi \in \Pi(A)$ such that the two atoms of $\pi$ are connected in $\mathcal G$.
    
        \begin{theorem}\label{thm:partition-for-A-2}
            Let $X_{[m]}$ be a Markov random field  with underlying connected graph $\mathcal G = ([m], E)$, and let $A = \{p, q\}$, $p \neq q$. For $B \subseteq [m]$, let $\mathcal{G}[B]$ denote the subgraph induced by $B$. Then, 
            \begin{align*}
                C_\mathrm{SK}(A) = \min_{\substack{B:\, p \in B,\, q \in B^c, \\ \mathcal{G}[B],\, \mathcal{G}[B^c] \text{ connected} }} I(X_B \wedge X_{B^c}).
            \end{align*}
        \end{theorem}
        
        \begin{proof}
            Let $(B, B^c) \in \Pi(A)$ be an optimal bipartition, and denote $\mathcal{I}(B)=I(X_B\wedge X_{B^c})$. If both induced subgraphs $\mathcal G [B]$ and $\mathcal G [B^c]$ are connected, there is nothing to prove. If not, at least one of $\mathcal G [B]$, $\mathcal G [B^c]$ must consist of at least two (maximally) connected components, which we call \emph{subatoms}. 
            At least one such subatom contains neither $p$ nor $q$, since there are at least three subatoms and only two of them can contain \(p\) or \(q\). Without loss of generality, let $B'\subsetneq B$ be such a subatom, and let $C = B \setminus B'$. 
    
            We construct the new bipartition $\{B\setminus B',\;B^c\cup B'\} = \{C,\; B^c \cup B'\}$, which is valid since neither $p$ nor $q$ belongs to $B'$. Since $B'$ is a subatom, there is no edge between $B'$ and $C$, and so $B^c$ must separate $B'$ and $C$. Using the global Markov property, 
            \begin{align}
                X_{B'} \mc X_{B^c} \mc X_C. \label{eq:2-part-proof-30} 
            \end{align} 
            The objective $\mathcal I(\cdot)$ does not increase, since
            $$\mathcal I(C)-\mathcal I(B) = I(X_C\wedge X_{B'}X_{B^c})-I(X_{B'}X_C\wedge X_{B^c}) = I(X_C\wedge X_{B^c})-I(X_{B'}X_C\wedge X_{B^c}) \leq 0 $$
        where the second equality follows from
        \eqref{eq:2-part-proof-30}. 
    
        This operation also strictly decreases the total number
        of connected components. Indeed, since \(\mathcal G\)
        is connected, \(B'\) has neighbors in at least one
        component of \(\mathcal G[B^c]\). If it has neighbors
        in \(k\geq1\) such components, then moving \(B'\)
        removes one component from \(\mathcal G[B]\) and
        merges \(B'\) with those \(k\) components into one.
        The total number of components therefore decreases
        by \(k\).
            
            Repeating this construction eventually leads to a valid bipartition $\{\bar{B},\bar{B}^{c}\}$ such that the total number of subatoms is $2$, i.e., both $\mathcal{G}[\bar{B}], \mathcal{G}[\bar{B}^c]$ are connected, and $\mathcal I (\bar B) \leq \mathcal I (B)$. If $(B, B^c)$ were an \emph{optimal} bipartition, then so is $(\bar B, \bar B^c)$, completing the proof.
        \end{proof}

        As a corollary, we recover the $|A|=2$ special case of the aforementioned generalization of  \Cref{eq:sk-trees} when $\mathcal G$ is a tree. The following special case for cycles is new. On a cycle, a bipartition separating \(p\) and \(q\) has both atoms connected exactly when its boundary consists of one edge from each of the two edge-disjoint \(p\)-to-\(q\) paths.
        \begin{corollary}\label{cor:A-2-trees}
            Let $X_{[m]}$ be a Markov random field on a cycle $\mathcal G=([m],E)$, and let
            $A=\{p,q\}$, $p\neq q$. Let $P(p,q)$ and $P'(p,q)$ denote the two edge-disjoint paths between $p$ and $q$ in $G$. Then
            \begin{align*}
                C_{\mathrm{SK}}(A) = \min_{e \in P(p,q),\, e' \in P'(p,q)} I(X_{C(e,e')}\wedge X_{C(e,e')^c}),
            \end{align*}
            where $C(e,e')$ is the vertex set of the component containing $p$ after deleting the edges $e, e'$.
        \end{corollary}
    
        \begin{proof}
            A bipartition of a cycle has both blocks connected
            exactly when it is obtained by deleting two distinct
            edges. It separates \(p\) and \(q\) exactly when
            one deleted edge lies on each of the two paths
            between them. The result therefore follows from
            \Cref{thm:partition-for-A-2}.
        \end{proof}
        On a cycle, the number of candidate bipartitions is therefore
        \[
            |P(p,q)|\,|P'(p,q)|
            \leq
            \left\lfloor\frac{m^2}{4}\right\rfloor,
        \]
        rather than the \(2^{m-2}\) bipartitions separating the two
        targets without the connectedness restriction.
    
    \section{Conclusion}\label{sec:conclusion}
        We have settled the question of tightness of the partition upper bound by Csisz\'ar--Narayan: the bound is tight
        for every source for \(m=2,3,4\), and \(5\),
        and any \(|A|\in \{2,3,\dots,m\}\), and for \(m\geq6\), it is tight when
        \(|A|\in\{2,m-2,m-1,m\}\) and not tight in general for any other
        \(|A|\). Our proofs used the properties of the linear program
        \Cref{eq:rco-lp} and the supermodularity of conditional entropy.
        We conclude with a few remarks.
        
        First, our results carry over to the \emph{private key capacity}
        (see \cite{csiszar2004secrecy}). In this setting, every terminal
        in \([m]\) sees, in addition to \(X_i^n\), the sequence \(Z^n\)
        for some \(Z\) jointly distributed with \(X_{[m]}\), and the
        generated key must be
        nearly independent of the joint observation consisting of
        the interactive public communication and \(Z^n\).
        The corresponding partition upper bound is the same as
        \Cref{eq:upper-bound}, except with additional conditioning on
        \(Z\) throughout. Since supermodularity is unaffected by this
        additional conditioning, our proofs go through verbatim. We did
        not include this in order to avoid
        notational clutter. The same counterexample as
        \Cref{sec:counterexample} works too, by simply taking \(Z\) to be
        an arbitrary discrete rv independent of \(X_{[m]}\).
        
        Second, for jointly Gaussian rvs with a positive-definite covariance
        matrix, an equality similar to \Cref{eq:si-rco-rel} holds, with
        differential entropies replacing entropies, except that the
        corresponding LP value cannot be interpreted as a communication
        rate \cite{nitinawarat2012gaussian}. Since conditional differential
        entropy is supermodular, with the convention
        \(h(X_\emptyset)=0\), our proofs in \Cref{sec:proofs} together with \Cref{rem:general-supermodular-objective}
        shows that the partition upper bound \Cref{eq:sk-capacity} holds with equality for $|A| \in \{2, m-2, m-1, m\}$ in the Gaussian setting without
        requiring strict supermodularity.
        This equality does \emph{not} hold in general for other values of $|A|$, though, since we can construct a counterexample\footnote{\samepage
        For example, let \(A=\{1,2,3\}\), and let
        \(U,V,W,N_1,\ldots,N_6\) be independent standard Gaussian rvs.
        Define
        \[
        \begin{aligned}
        X_1&=U+W+N_1, & X_2&=V+W+N_2, & X_3&=U-V+N_3,\\
        X_4&=U+N_4, & X_5&=V+N_5, & X_6&=W+N_6. 
        \end{aligned}
        \]
        
        For this source, the Gaussian analogue of the
        partition upper bound equals \(\frac14\log_2\frac{12}{5}\),
        whereas the exact secret key capacity is $C_{\mathrm{SK}}(A)
            =
            \frac18\log_2\frac{3456}{625}
            <
            \frac14\log_2\frac{12}{5}.$} with $m=6$ and $|A| = 3$.
        A construction similar to the proof of \Cref{prop:counterexample-extension} for other pairs $(m,|A|)$ fails because the resulting covariance matrix is not positive-definite.
        
        Finally, when \(|A|=m\), the optimization \Cref{eq:sk-capacity}
        has several appealing combinatorial properties, particularly the
        existence of a unique finest partition, which leads, for example,
        to algorithms for successive omniscience
        \cite{chan2016successive}.
        An interesting question is to identify conditions under which
        analogous structural properties, and their consequences, continue
        to hold when \(|A|\neq m\).

\printbibliography 
\end{document}